\documentclass[a4paper,12pt]{article}
\pdfoutput=1
\usepackage[utf8]{inputenc}
\usepackage[T1]{fontenc}
\usepackage{hyperref}
\usepackage{epsfig}
\usepackage{amssymb}
\usepackage{color}
\usepackage{xcolor}
\usepackage{setspace}

\usepackage{amsmath}
\usepackage{amssymb}
\usepackage{geometry}
\usepackage{subfigure}
\usepackage{url}
\usepackage{slashed}
\usepackage{booktabs}%
\usepackage{float}
\usepackage{graphicx}

\usepackage{pdflscape}

\newcommand{\be}{\begin{equation}}
\newcommand{\ee}{\end{equation}}
\newcommand{\br}{\begin{eqnarray}}
\newcommand{\bea}{\begin{eqnarray}}
\newcommand{\eea}{\end{eqnarray}}
\newcommand{\er}{\end{eqnarray}}
\newcommand{\ba}{\begin{array}}
\newcommand{\ea}{\end{array}}
\newcommand{\bi}{\begin{itemize}}
\newcommand{\ei}{\end{itemize}}
\newcommand{\bn}{\begin{enumerate}}
\newcommand{\en}{\end{enumerate}}
\newcommand{\bc}{\begin{center}}
\newcommand{\ec}{\end{center}}

\newcommand{\gsim}{\lower1.0ex\hbox{$\;\stackrel{\textstyle>}{\sim}\;$}}
\newcommand{\lsim}{\lower1.0ex\hbox{$\;\stackrel{\textstyle<}{\sim}\;$}}

\newcommand{\bs}{\begin{small}}
\newcommand{\es}{\end{small}}

\begin{document}
\pagestyle{empty}
\begin{flushright}
CP3-20-15\\  LAPTH-013/20\\  MCNET-20-11\\
\end{flushright}
\vspace*{10mm}
\begin{center}
{\LARGE {\bf 
Measuring the quartic Higgs self-coupling 
\\
\vspace{0.5cm}
at a multi-TeV muon collider
}} \\
\vspace*{1.5cm}

{\bf Mauro Chiesa$^{{a}}$, Fabio Maltoni$^{{b,c}}$, Luca Mantani$^{{b,d}}$, Barbara Mele$^{{e}}$, \\[5pt]
Fulvio Piccinini$^{{f}}$, Xiaoran Zhao$^{{b}}$}\\
\vspace{0.5cm}
{\it
(a) CNRS, LAPTh, 9 Chemin de Bellevue, 74940 Annecy, France\\
(b)Centre for Cosmology, Particle Physics and Phenomenology {\rm (CP3)},\\
Universit\'e Catholique de Louvain, Chemin du Cyclotron, B-1348 Louvain la Neuve, Belgium\\
(c) Dipartimento di Fisica e Astronomia, Universit\`a di Bologna e INFN, Sezione di Bologna, via Irnerio 46, I-40126 Bologna, Italy \\
(d) Institut f\"ur Theoretische Physik, Universit\"at Heidelberg, Germany\\
(e) INFN, Sezione di Roma, c/o Dipartimento di Fisica, "Sapienza" Universit\`a di Roma, P.le Aldo Moro 2, I-00185 Rome, Italy\\
(f) INFN, Sezione di Pavia, Via A. Bassi 6, I-27100 Pavia, Italy}

\vspace*{2cm}{\bf ABSTRACT}
\end{center}

\vspace{0.3cm}

Measuring the shape of the Higgs boson potential is of paramount importance, and will be a challenging task at
current as well as future colliders. While the expectations for the measurement of the trilinear Higgs self-coupling are
rather promising, an accurate measurement of the quartic self-coupling interaction is presently considered extremely
challenging even at a future 100 TeV proton-proton collider. 
In this work we explore the sensitivity that a muon collider with a center of mass energy in the multi-TeV range 
and luminosities of the order of $10^{35}$cm$^{-2}$s$^{-1}$, as presently under discussion, might provide, thanks to
a rather large three Higgs-boson production and to a limited background.  By performing a first and simple analysis,
we find a clear indication that a muon collider could provide a determination of the quartic Higgs self-coupling that
is significantly better than what is currently considered attainable at other future colliders.

\vspace*{5mm}

\noindent


\vfill\eject

\pagestyle{plain}


\section{Introduction}
The Higgs boson discovery at the LHC in 2012~\cite{Aad:2012tfa,Chatrchyan:2012ufa} and the subsequent campaign of measurements of its properties~\cite{Sirunyan:2018koj,Aad:2019mbh}, have provided a wonderful confirmation of our understanding of elementary particles and their interactions at the weak scale. So far, the predictions of the Standard Model (SM) for the Higgs boson couplings to the vector bosons and to third generation fermions are in spectacular agreement with observations~\cite{Englert:1964et,Higgs:1964ia,Higgs:1964pj,Guralnik:1964eu}. On the other hand, its interactions with lighter sectors, such as the first and second generation quarks and leptons,  are still to be confirmed. In addition, the very existence of a scalar doublet has opened many possibilities for new physics to couple to the SM as well as many new avenues for searching for it. Leading, yet simple examples are Higgs portals to singlet fermions and/or scalars, which could provide a solution to open questions such as that of the nature of dark matter or the origin of matter-anti-matter asymmetry in the universe. Other possibilities could involve extended charged scalar sectors, which in turn could point to the existence of new symmetries, such as supersymmetry.  
All such possibilities are continuously pushed at higher scales by the accuracy of the measurements of the  Higgs-boson couplings to the heavier SM particles, which is presently O(10\%)~\cite{Sirunyan:2018koj,Aad:2019mbh}. The high luminosity phase (HL-LHC) will improve the corresponding accuracy to a few percents for the vector bosons and third generation, and access for the first time the couplings to the second generation fermions~\cite{Cepeda:2019klc,deBlas:2019rxi}.  

One key sector, which is currently very weakly constrained and could very easily hide or be connected to new physics, is the scalar potential~\cite{deBlas:2019rxi,Agrawal:2019bpm}.
In the SM, the Higgs scalar potential is fixed by just two low energy parameters, the Higgs mass ($m_H\simeq 125\!$~GeV) and the Fermi constant $G_F$ (or equivalently the vacuum expectation value $v\simeq 246\!$~GeV). At the weak scale, the potential can be written in terms of the Higgs  trilinear ($\lambda_3$) and quartic ($\lambda_4$) self-couplings
\begin{equation}
V(H)=\frac{1}{2}m_H^2H^2 + \lambda_3 v H^3 + \frac{1}{4} \lambda_4 H^4,
\end{equation}
where in the SM,  $\lambda_3=\lambda_4=m_H^2/2v^2\equiv\lambda_{SM}$. In particular, higher-point Higgs boson self interactions are forbidden by the request of renormalisability of the SM up to arbitrarily high scales. 

The measurement of the parameters that describe the shape of the Higgs potential  are therefore a milestone in the quest of understanding the mechanism of the electroweak symmetry breaking and of exploration of new physics. The relevance of this information on the one hand,  and the inherent experimental challenges on the other hand, make this measurement one of the most relevant benchmarks that can be employed to set the physics potential of future high-energy collider projects. 

The determination of the trilinear Higgs self-couplings $\lambda_3$ at the LHC
and at future colliders has been vastly considered in the literature, from measurements involving Higgs boson pair production and through radiative effects in single Higgs production (see for instance~\cite{deBlas:2019rxi} and~\cite{DiMicco:2019ngk}).
At the end of the complete FCC programme~\cite{Abada:2019lih,Abada:2019zxq,Benedikt:2018csr} 
one expects to reach a ${\cal O}(5\%)$ accuracy on $\lambda_3$~\cite{Abada:2019lih}.

Although in SM extensions where new physics is at higher scales, $\lambda_4$ is related to $\lambda_3$, 
an unbiased determination of the Higgs quartic self-coupling will require a measurement
of processes genuinely depending on $\lambda_4$, like the triple Higgs production.\footnote{Double Higgs production is sensitive to $\lambda_4$ through loop effects, see \cite{Maltoni:2018ttu,Bizon:2018syu,Borowka:2018pxx}.}

The measurement of the triple Higgs production cross section, currently being the most  studied 
handle on the quartic Higgs self-coupling, looks very challenging even at the 100 TeV
proton collisions foreseen at the FCC-hh. Quite a number of studies concerning different  final states deriving from various combinations of the Higgs decay channels have been considered, see~\cite{DiMicco:2019ngk}. The expected constraint on a $\lambda_4$ deviation (for a SM value of $\lambda_3$) is quite poor~\cite{Papaefstathiou:2015paa,Contino:2016spe,Fuks:2017zkg}, the most optimistic estimate obtained from $HHH$ production with $6b$ in the final state is $\lambda_4/\lambda_4^{\rm SM} \in $ [-2,+13] (at 2$\sigma$, with $\lambda_3=\lambda_3^{\rm SM}$) with a significance for SM $HHH$ production of about $2\sigma$ with 20 ab$^{-1}$ of integrated luminosity (and perfect $b$-tagging)~\cite{Papaefstathiou:2019ofh}.
Indirect bounds on the quartic Higgs self-coupling can be obtained from one-loop contributions in $HH$ final states
at future lepton and hadron colliders (see Refs.~\cite{DiMicco:2019ngk,Maltoni:2018ttu,Bizon:2018syu,Borowka:2018pxx} and references 
therein) and these contributions allow to constrain $\lambda_4/\lambda_4^{\rm SM}$ at FCC-hh in the range [-2.3,+4.3]
at 1$\sigma$ for $\lambda_3=\lambda_3^{\rm SM}$~\cite{DiMicco:2019ngk}.

The aim of the present study is to explore for the first time the reach of a multi-TeV muon collider for a
complete reconstruction of the shape of the Higgs potential. 
In connection with the discussion on next generation high-energy colliders
carried out for the 2020 European Strategy Update on Particle 
Physics~\cite{Strategy:2019vxc}, a very attractive option was given by the possibility of 
a high-luminosity multi-TeV muon collider~\cite{Delahaye:2019omf}. In particular,
a collider with center of mass (c.m.) energies in the range 1.5 to 14 TeV,
and luminosities up to ${\cal O}(10^{35})$~cm$^{-2}$s$^{-1}$ is presently under consideration. 
Although a long and challenging period of further accelerator research and development is still needed to prove the actual feasibility of such a machine, its physics opportunities seem extremely  wide and rich and therefore need to be carefully assessed.  

There are a number of immediate and crucial advantages in replacing electrons with muons in lepton collisions,
that would allow to amazingly extend the effective collision energy in realistic colliders.
For instance, in the LHC tunnel, one might implement 14 TeV muon collisions~\cite{Shiltsev:2019rfl},
whose discovery potential in direct searches of heavy (SM charged)
states would be roughly similar to the one of a 100 TeV proton collider of similar luminosity~\cite{Delahaye:2019omf}. 
In addition, accelerating  muons  could offer a very cost-effective way to increase
the lepton collision energy reach, while keeping the beam energy spread
one order of magnitude smaller than for an electron collider of 
similar c.m. energy~\cite{Shiltsev:2019clx}. Finally,  progress on long-standing hurdles has been recently achieved. For example,  preliminary studies show that potentially serious beam-induced 
background effects arising from the beam muon decays could be manageable as they become
less severe at higher c.m. energies~\cite{Bartosik:2019dzq,Lucchesi}. Nevertheless, collecting data in the forward regions  in such a high-background environment might be challenging, and  need more ideas and developments~\cite{Foster:1995ru, Johnstone:1996hp, Mokhov:2011zzd, PhysRevSTAB.14.061001, Mokhov:2014hza}.

In the following, we assume four hypothetical setups for the 
c.m. energy and  luminosity as references:
$\sqrt{s} \simeq [1.5,3,6,14]\!$~TeV and ${\cal L}\simeq [1.2,4.4,12,33]\cdot 10^{34}$cm$^{-2}$s$^{-1}$, respectively.
These configurations are  based  on the parameters characterizing present  muon collider designs  according to the MAP scheme~\cite{Shiltsev:2019clx,Palmer,Schulte,Shiltsev:2019rfl}.~\footnote{For machine designs using resonant production of muon pairs
  at threshold from e+e- collisions, see~\cite{Antonelli:2015nla,Antonelli:2019uoe}.}
  In addition, we will consider two  further collision energies/luminosities, i.e. $\sqrt{s}\simeq [10, 30]\!$~TeV and ${\cal L}\simeq [20,100]\cdot10^{34}$cm$^{-2}$s$^{-1}$, respectively, motivated by the required scaling of the luminosity  needed to compensate the $1/s$ decrease in the $s$-channel cross sections that are relevant for pair production of new heavy objects~\cite{Delahaye:2019omf}.
The setups are summarized in  table~\ref{tab:sL}, where for each $\sqrt{s}$ value  we also report  the integrated luminosity (L) collected over  a ten-year run  (with a conventional year of $10^7$ seconds).
\begin{table}
 \begin{center}  
  \begin{tabular}{| l || c || c || c || c || c || c || c || c || c |}
      \hline
    $\sqrt{s}$ \;\;\;\; \;\;\;\;(TeV)         & 1.5   &   3    & 6       & 10     & 14      & 30    \\
  \hline\hline
      ${\cal L}$\;\;\; ($10^{34}$\,cm$^{-2}$s$^{-1}$)        & 1.2  &   4.4  & 12    & 20    & 33 & 100    \\
       \hline\hline
      L$_{10y}$\; \;\;\;\; \;(ab$^{-1}\!)$        & 1.2  &   4.4  & 12    & 20    & 33 & 100    \\
  \hline
\end{tabular}
\caption{\label{tab:sL}  Reference muon  collision energies $\sqrt s$, and  instantaneous luminosities ${\cal L}$, with corresponding integrated luminosities L for a 10 years run (one year of $\sim10^7$s). The luminosity values assumed for $\sqrt{s}\simeq (1.5,3,6,14)$~TeV are as
from \cite{Shiltsev:2019clx,Schulte}.
}
\end{center}
\end{table}

A high-luminosity multi-TeV muon collider has a physics potential both for direct searches of heavy objects as well as for precision measurements~\cite{Delahaye:2019omf}. As a prime example of the latter, in this work, we focus on its capability to constrain the SM Higgs scalar potential.
The reach of such a measurement builds up on the clean environment of lepton collisions, where QCD backgrounds
are moderate, which also allows events to be recorded in absence of a trigger.  
A few percent determination of  the trilinear Higgs self-coupling $\lambda_3$ via double Higgs production, eventually even better than that achievable at the FCC, seems possible at the moment~\cite{Delahaye:2019omf}, although only simplified studies are available~\cite{Conway:2014gaa}.

In this paper, we provide  a first quantitative analysis of the muon collider potential to access information on the {\it quartic} Higgs self-coupling $\lambda_4$ as obtained from direct measurements of the cross section for  triple Higgs-boson  production.
We will consider in particular the multi-TeV energy and order 10$^{35}$ cm$^{-2}$s$^{-1}$ luminosity options  considered in  table~\ref{tab:sL}.  
Since, for $\sqrt{s} \gtrsim 1.5$ TeV, vector-boson-fusion channels 
(whose cross sections grow as $\log s$)  get the upper-hand on the corresponding $s$-channel production mediated by the $\mu^+ \mu^- \to H H H Z^\ast$ process,
our analysis will be mainly focused on the W-boson-fusion (WBF) process
\begin{equation}
\mu^+ \mu^- \to W^\ast W^\ast \nu_\mu \overline{\nu}_\mu\to H H H \nu_\mu \overline{\nu}_\mu .
\end{equation}
Depending on the particular Higgs decay channel involved, the final signature 
of triple Higgs events can be quite diverse~\cite{Papaefstathiou:2015paa}, although a few kinematical common features
(like the presence of three systems resonating at the Higgs masses) will be universal.
Thanks to these features, even dijet final states, such as the 
$b$-jets from high-rate $H\to b\bar b$ decays, are expected to be efficiently reconstructed.~\footnote{In fact,  the final detection efficiency could be strongly affected by the 
machine-induced background and the machine detector interface that could seriously impact  the final detector acceptance~\cite{Bartosik:2019dzq,Lucchesi}.
In any case, it is clear that further research and development of accelerator, detector, and analysis technologies for a multi-TeV muon collider will be needed to reach robust conclusions on the physics potential of such a machine. }

In this study, we work under two main hypotheses.  First,  we assume that a number of potential machine and detector issues will be  solved after detailed studies, possibly involving innovative technologies, and 
discuss  the potential consequences of just having at disposal signal event statistics for triple Higgs bosons   
corresponding to such high c.m. energies and luminosities as envisaged in the MAP project. 
Second, we assume that the bulk of the different $HHH$ final states corresponding to the dominant Higgs decay channels can be reconstructed  with high efficiency. Correspondingly, we estimate the muon collider sensitivity
to detect a deviation in the Higgs $\lambda_3$ and $\lambda_4$ self-couplings through the full statistics of the triple Higgs
production. On  the other hand, as far as the Higgs trilinear self-coupling $\lambda_3$ is concerned, we do not consider here the  stronger direct constraints  that presumably can be obtained through the scrutiny of the higher-rate double Higgs production. 

The plan of the paper is as follows. In Section 2 we present the results of our Monte Carlo simulations for the signal cross sections and distributions in the standard model, including a study of the acceptance of $6$ b-jet final state.

In Section 3, we parametrise the cross sections dependence in new physics scenarios as a quartic polynomial of the deviations $\delta_3$ and  $\delta_4$ of the self-couplings with respect to the SM predictions and study the sensitivity of representative distributions to them. Finally, we determine the constraining potential (considering different energy and luminosity setups)  of a future muon collider. In the last section, we present our conclusions and  the outlook.

\section{Triple Higgs production in the standard model}

In this section, we present the cross sections and a few kinematical distributions for the process
\begin{equation}
\mu^+ \mu^- \to H H H \,\nu \overline{\nu} ,
\label{HHH}
\end{equation}
in the SM and in scenarios where the Higgs self-couplings are modified, at muon collider energies in the range [1.5, 30] TeV. 

\begin{figure}
\begin{center}
  \includegraphics[scale=0.5]{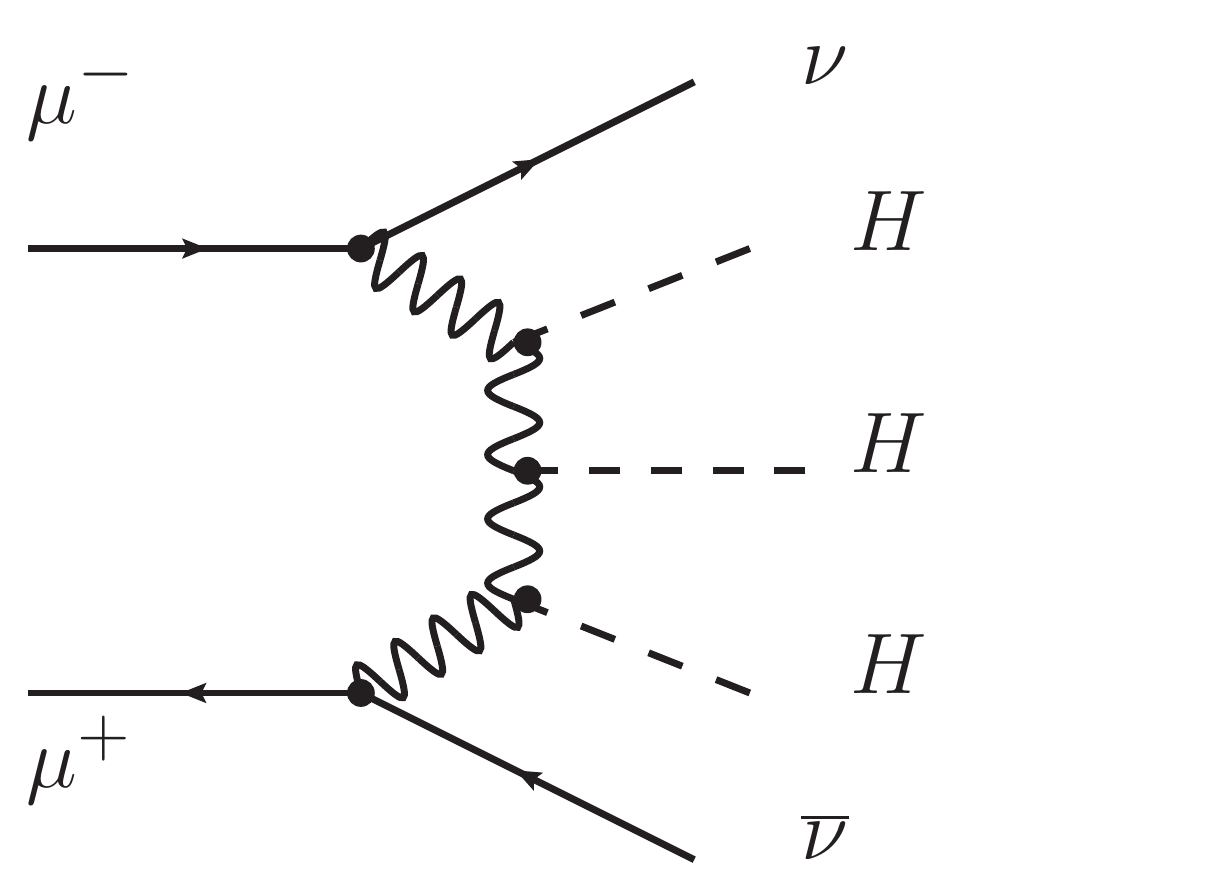}   \includegraphics[scale=0.5]{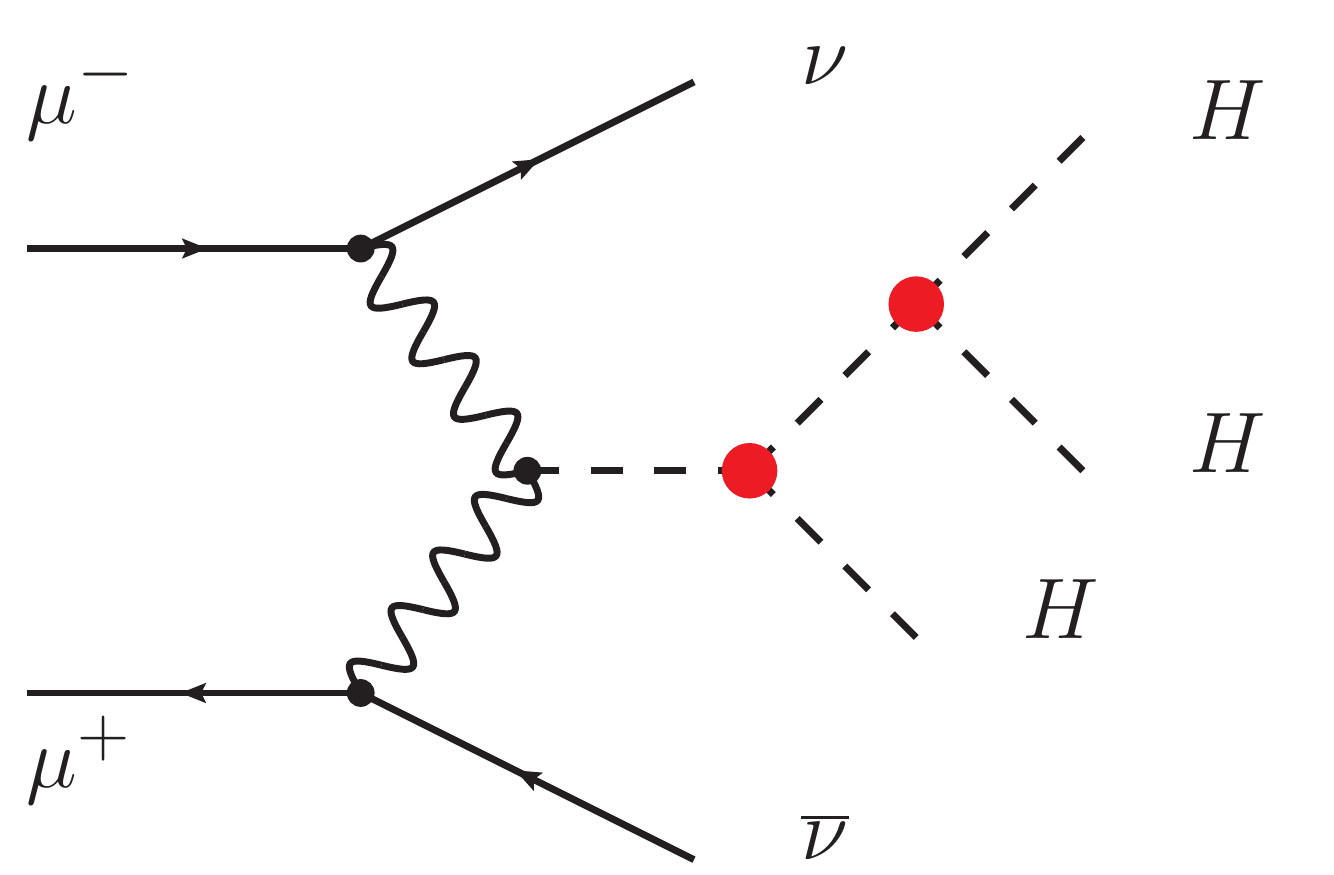}    
  \vskip0.5cm
  \includegraphics[scale=0.5]{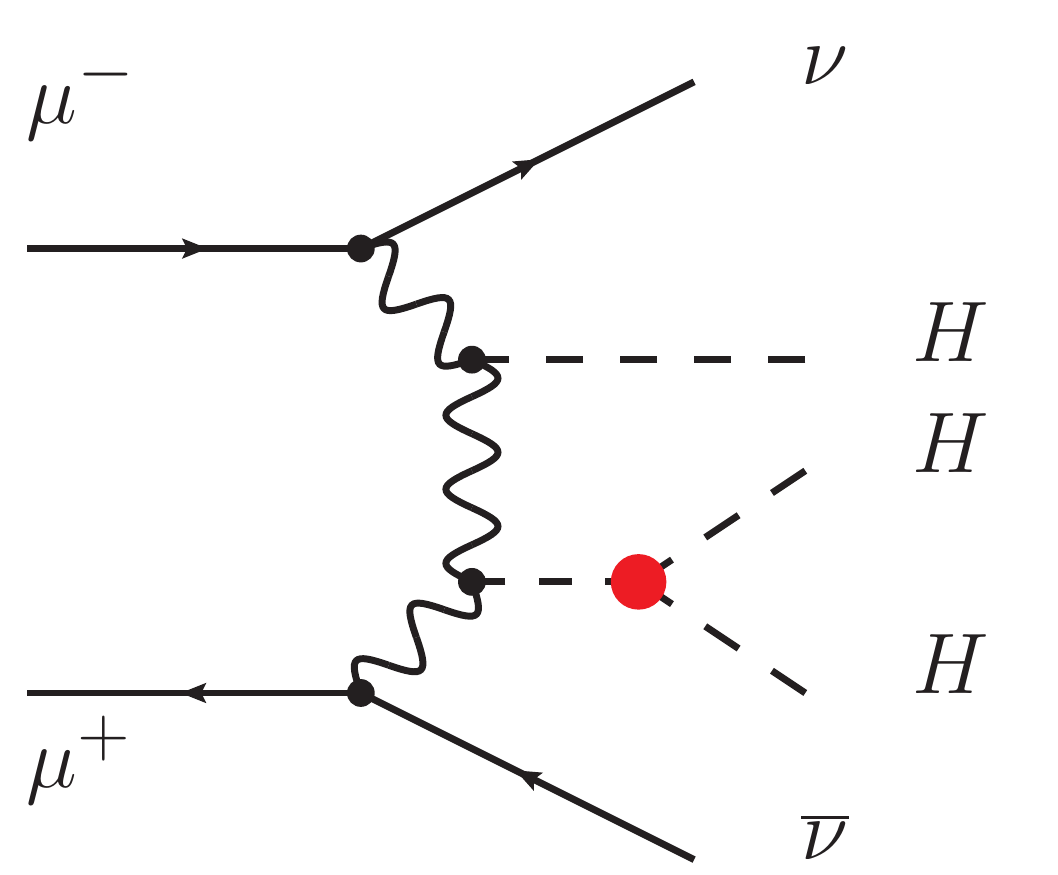}   \includegraphics[scale=0.5]{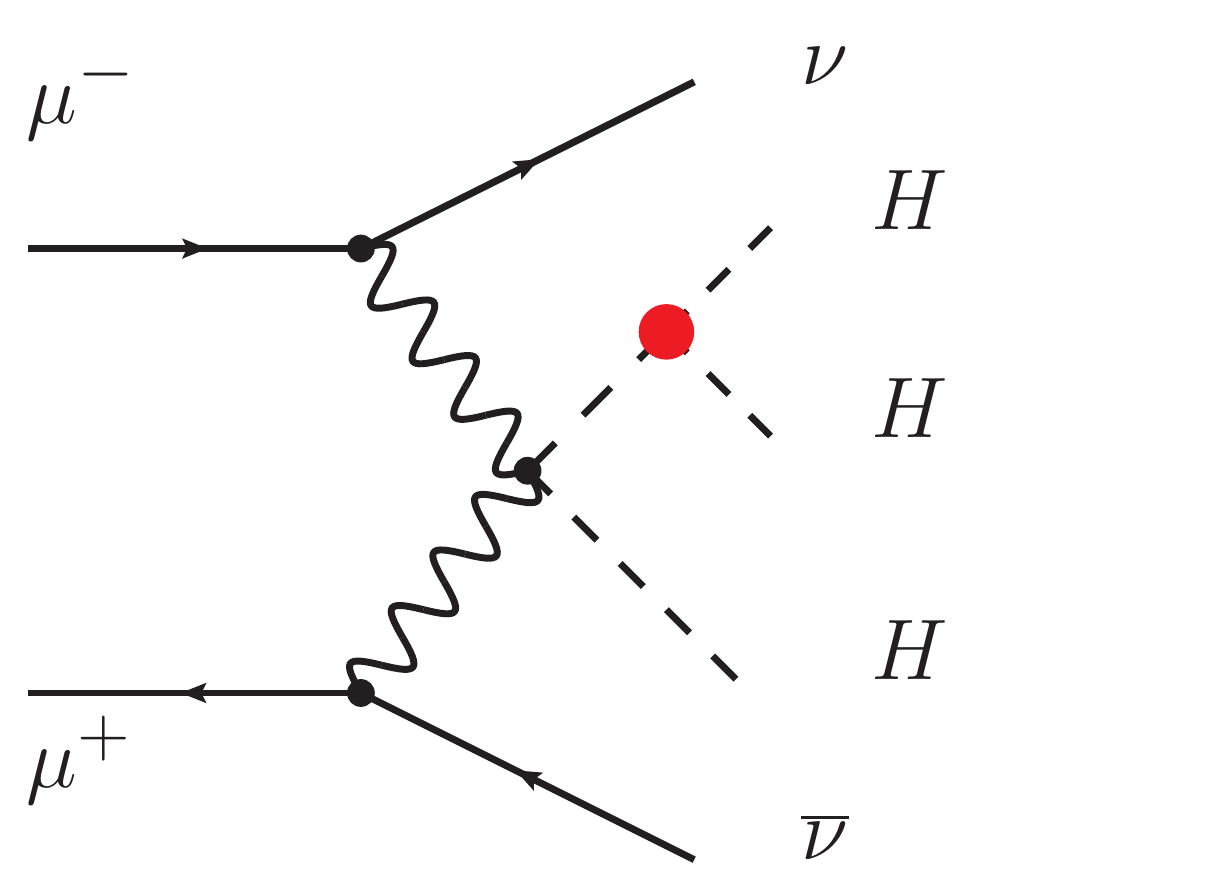}
  \vskip0.5cm
  \includegraphics[scale=0.5]{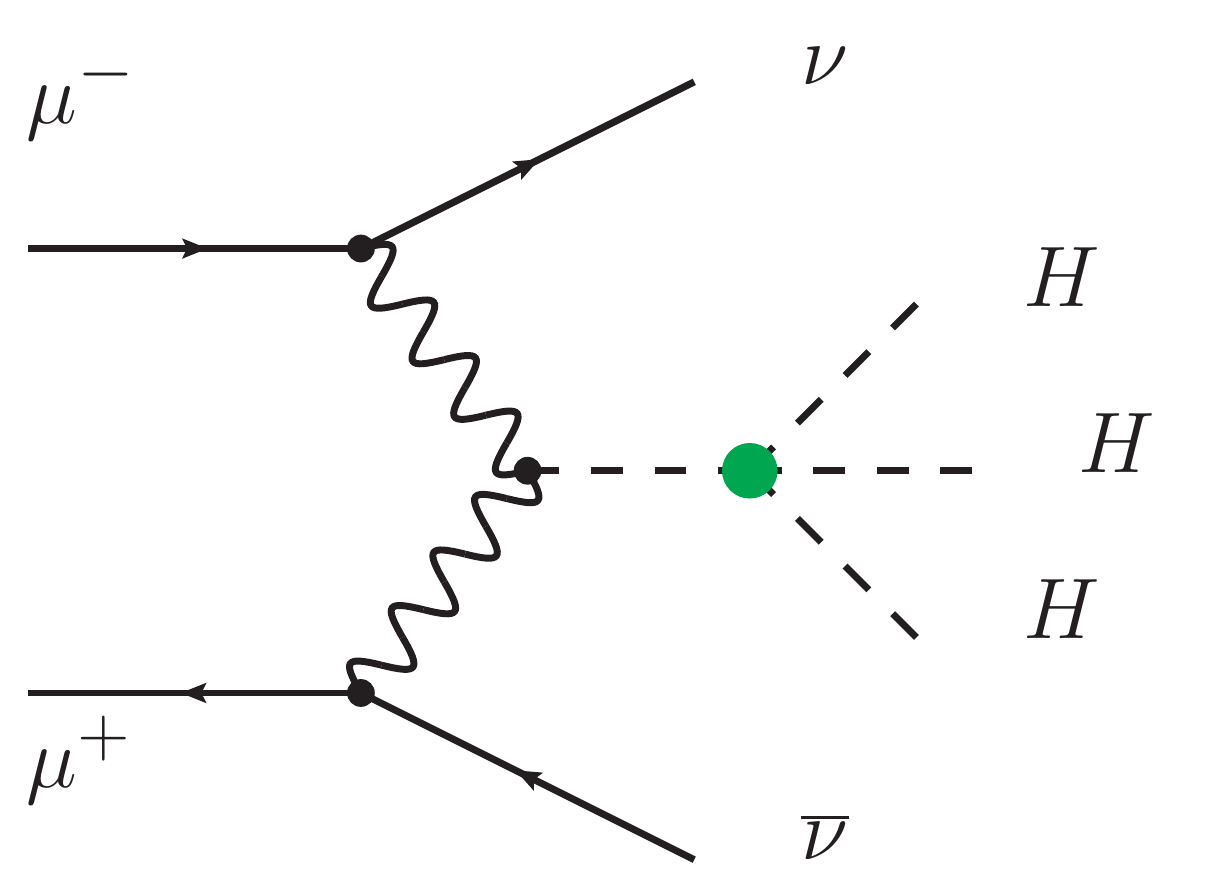}    \includegraphics[scale=0.5]{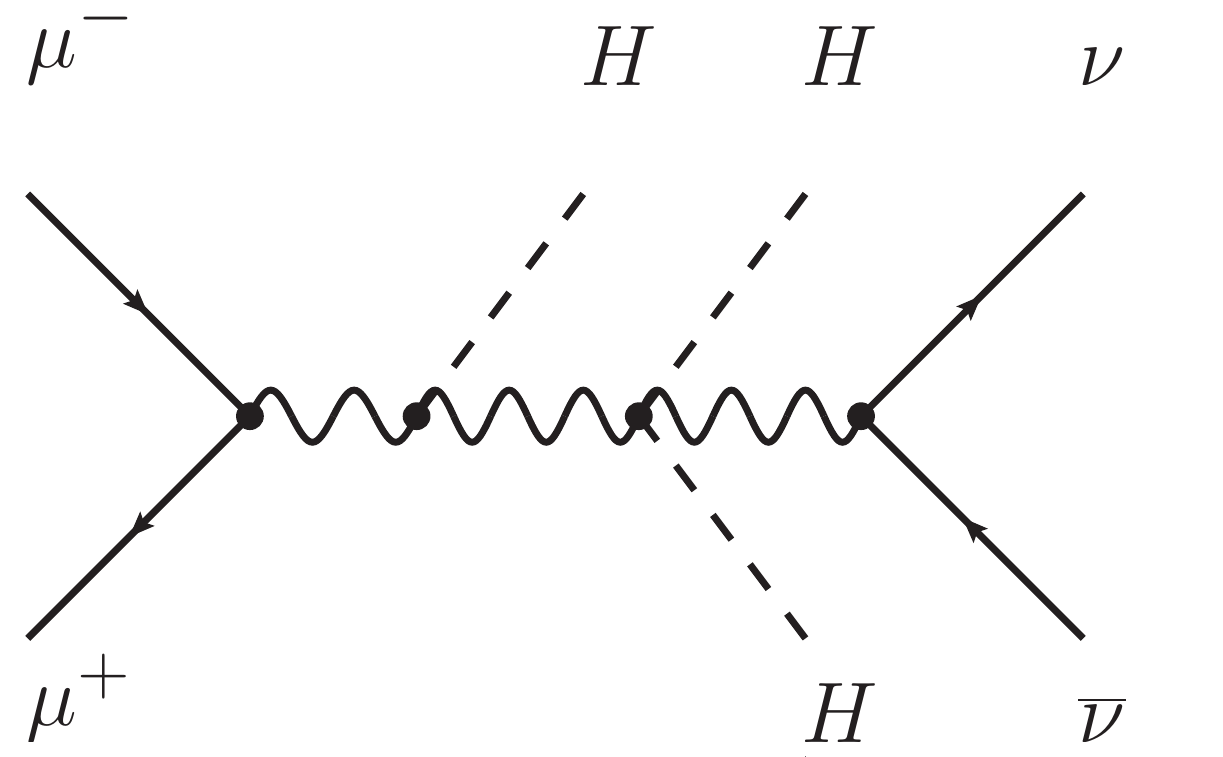}
  \caption{\label{Feyndiags} Representative Feynman diagrams contributing to the process
    $\mu^+ \mu^- \to H H H \nu \overline{\nu}$ that do not involve self-couplings (top-left and bottom-right), involve the trilinear twice (top-right) and once (central), and the quartic (bottom-left)  couplings. $s$-channel diagrams (bottom-right) contribute but become negligible at high energy (note that in this case $\nu=\nu_e,\nu_\mu,\nu_\tau$). }
    \end{center}
\end{figure}

In figure~\ref{Feyndiags}, we show  a few representative Feynman diagrams of the process.
By inspection, one can quickly conclude that at the tree level, each diagram can be at most linearly dependent on
the quartic self-coupling $\lambda_4$, and linearly or quadratically dependent on~$\lambda_3$.
In fact, the majority of diagrams are independent from Higgs self-couplings. 
This observation leads to the expectation that on the one hand,
the cross section sensitivity to self-couplings in general and to the quartic coupling in particular,  
will be quite mild and on the other hand, a very precise knowledge of the $WWH$ and $WWHH$ couplings will be needed in order to pin down the Higgs potential.  

Triple Higgs production proceeds through two main classes of diagrams:  the WBF channel~\footnote{The corresponding cross sections for $Z$ boson fusion, $\mu^+ \mu^- \to Z^\ast Z^\ast \mu^+ \mu^- \to H H H\, \mu^+ \mu^-$ amount to 15--20\% of the ones for $W$ boson fusion, and are therefore relevant. We leave their inclusion to future work.  }
\begin{equation}
\mu^+ \mu^- \to W^\ast W^\ast \nu_\mu \overline{\nu}_\mu\to H H H\, \nu_\mu \overline{\nu}_\mu ,
\end{equation}
and  the $s$-channel
\begin{equation}
\mu^+ \mu^- \to H H H \;Z^\ast \to H H H \,\nu_{e,\mu,\tau} \overline{\nu}_{e,\mu,\tau}\; .
\end{equation}
Both sets contribute at the amplitude level to  $\mu^+ \mu^- \to H H H \nu \overline{\nu}$ yet, as we will discuss in the following, mostly in different phase space regions.

In order to compute the $\mu^+ \mu^- \to H H H \nu \overline{\nu}$ cross sections and distributions, including the
complete self-coupling dependence, we have used two Monte Carlo
event generators: {\sc Whizard} \cite{Moretti:2001zz,Kilian:2007gr} (version 2.6.4) and  {\sc MadGraph5\_aMC@NLO} \cite{Alwall:2014hca}. 
Even though the SM implementation in both codes does not allow the user to change $\lambda_3$ and $\lambda_4$ from the input cards, it is sufficiently easy to do that directly accessing the source codes.~\footnote{In {\sc MadGraph5\_aMC@NLO} is also possible to use the {\sc SMEFT@NLO} model.}  The Higgs and gauge boson widths as well as the muon mass (and Yukawa) are set to zero, in order to avoid issues with gauge cancellations at very high energy.  

For all the results discussed in the following, we impose a technical generation cut  $M_{\nu \overline{\nu}}>150$~GeV 
on the neutrino pair invariant mass $M_{\nu \overline{\nu}}$, 
to prevent the singularity arising from a vanishing $Z$-boson width  
in the $s$-channel. The latter cut effectively takes away most of the $s$-channel 
contribution.
The  $s$-channel effects tend anyhow to be strongly suppressed at multi-TeV collision energies.
After removing the $Z$-resonance contributions by the  $M_{\nu \overline{\nu}}>150$~GeV cut, we find that the relative off-shell contribution of the
$\mu^+ \mu^- \!\!\!\to H H H Z^\ast\!\!\to H H H \,\nu \overline{\nu}$ to the total cross section is about 2.5$\%$ at $\sqrt s \simeq$1.5 TeV, 1.4$\cdot 10^{-3}$ at 3 TeV, and 10$^{-4}$ at 6 TeV.  With the present LO accuracy,  our complete results for 
$\mu^+ \mu^- \to H H H \,\nu \overline{\nu}$ will then match the ones for the WBF process
$\mu^+ \mu^- \to W^\ast W^\ast \nu_\mu \overline{\nu}_\mu\to H H H \nu_\mu \overline{\nu}_\mu$ with excellent accuracy in the energy range considered.~\footnote{Note that interference effects between the WBF and $s$-channel diagrams are  negligible due to the non-overlapping typical kinematics of the two configurations. For the reasons above,  in {\sc MadGraph5\_aMC@NLO} we find it easier to directly exclude the $s$-channel contributions by actually simulating $e^+ \mu^- \to H H H \nu_\mu \overline{\nu}_e$. We have explicitly checked that this approximation is excellent and make the simulations faster. }

\begin{figure}
  \begin{center}  
  \includegraphics[width=.45\columnwidth]{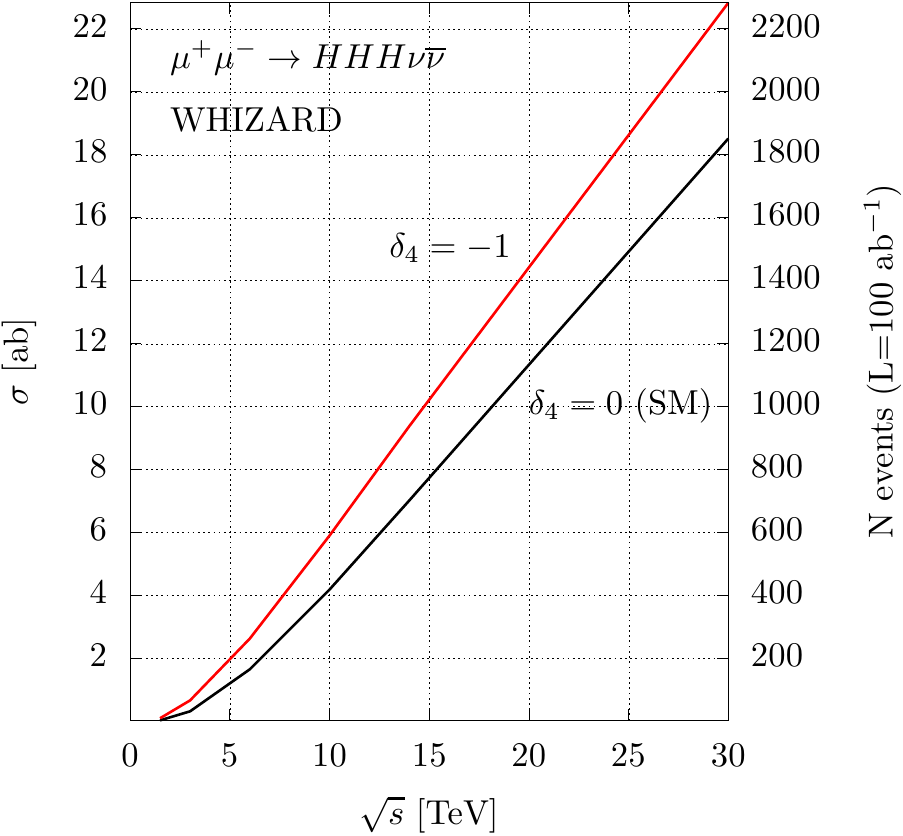}
  \includegraphics[width=.45\columnwidth]{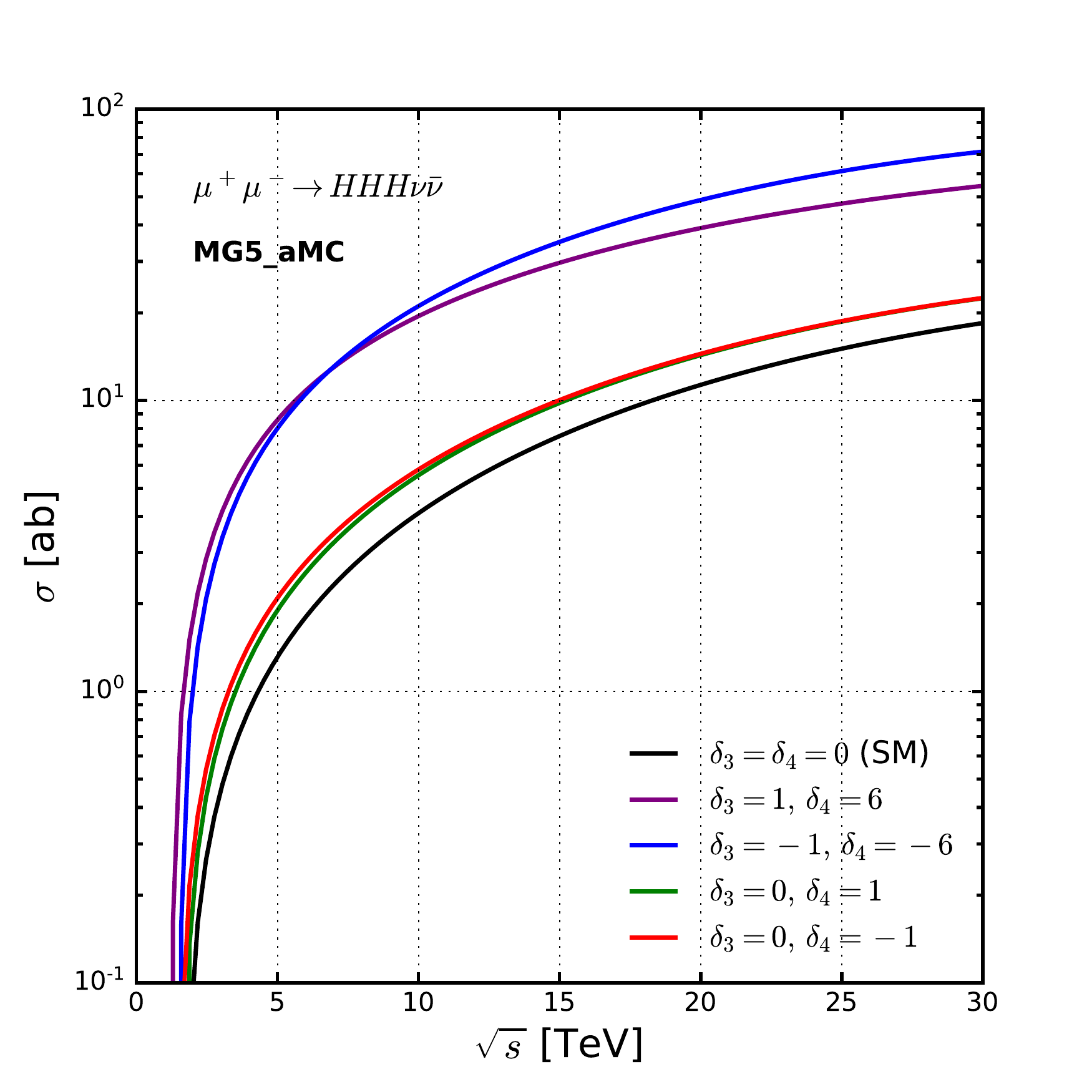}
  \caption{\label{total}   
  Expected cross sections (left) and signal event numbers  for a reference integrated luminosity of 100 ab$^{-1}$ (right) for $\mu^+ \mu^- \to H H H \nu \overline{\nu}$ versus the c.m. collision energy, for $M_{\bar\nu\nu}\gtrsim 150$GeV. 
 Cross sections for different assumptions of the trilinear and quartic couplings are presented, as well as for the SM case, obtained by {\sc Whizard} (left-hand side) and {\sc MadGraph5\_aMC@NLO} (right-hand side). Details on the scenarios are given in the text.}
 \end{center}
\end{figure}

In figure~\ref{total}, we plot the $\mu^+ \mu^- \to H H H \nu \overline{\nu}$ cross section versus $\sqrt s$ in the SM. On the right axis we include the expected number of triple Higgs final states produced for an integrated luminosity  L=100~ab$^{-1}$. The left-hand plot corresponds to the cross-sections results in a linear scale for two anomalous scenarios as obtained in {\sc Whizard}, while on the right-hand side the {\sc MadGraph5\_aMC@NLO} results for the yield are plotted in a log-scale, also for two additional scenarios. We have carefully verified that the results from the two MC's agree within uncertainties for SM as well as in presence of anomalous interactions.  We define $\delta_{3,4}$ and $\kappa_{3,4}$, through the following relations
\begin{eqnarray}
&&\lambda_3 = \lambda_{SM} ( 1 + \delta_3 ) = \kappa_3  \lambda_{SM} \,, \\
&&\lambda_4 = \lambda_{SM} ( 1 + \delta_4 ) = \kappa_4  \lambda_{SM}\,,
\end{eqnarray}
which imply that the SM values for the couplings are recovered for $\delta_{3,4}=0$, or equivalently for $\kappa_{3,4}=1$.
We point out that, for the sake of both simplicity and generality, we phrase our results in terms of 
the anomalous couplings above. At the perturbative level of our predictions, i.e., at the tree level, one can easily link the deformations of the $\lambda$'s to the coefficients of higher dimensional operators, see for instance~\cite{Borowka:2018pxx}. The simplest instance is that of adding just one operator of dimension six, $c_6 (\Phi^\dagger \Phi)^3/\Lambda^2 $. In this case, one finds that the shifts in the trilinear and quartic couplings are related, i.e., 
\begin{equation}
\delta_4= 6 \,\delta_3  , \, \qquad {( \rm SMEFT\;at\dim=6).}
\end{equation}
This constraint can be lifted by further adding operators of higher dimension, i.e., $c_8 (\Phi^\dagger \Phi)^4/\Lambda^4$. As special case of the latter situation, one can fix the couplings of the six and eight dimensional operators, to only have the quartic coupling modified, $\delta_3=0$ and $\delta_4\neq 0$. However, it is important to remind that this is not what is generically expected from the SMEFT and it implies a fine tuning, which is valid only at a given scale.

In order to get a first feeling of the cross section sensitivity to variations of the Higgs quartic coupling, 
in figures~\ref{total} we also show the cross section obtained by keeping the SM value for $\lambda_3$ and switching off $\lambda_4$ ($\delta_{3}=0, \delta_{4}=-1$ or $\kappa_{3}=1, \kappa_{4}=0$).  The effect is an increase, as expected from general arguments on unitarity cancellation, of production rates of about 20\%$-$30\% in the $\sqrt s$ range considered here.  On the right-hand plot, we show the corresponding results as obtained from {\sc MG5aMC} also including two scenarios of interest:  the $\delta_{3}=\pm 1, \delta_{4}=\pm 6$ cases, corresponding to  relative shift between $\delta_{3}$ and $\delta_{4}$ consistent with an EFT approach, and a scenario $\delta_{3}=0, \delta_{4}=+1$ with no change in  $\lambda_3$, yet a 100\% increase of $\lambda_4$. It is interesting to note that, as far as total rates are concerned, the latter case turns out to be  hardly distinguishable from the scenario where $\lambda_3=\lambda_{SM}$ and $\lambda_4=0$.
 
\begin{table}
 \begin{center}  
  \begin{tabular}{| c || c || c || c || c || c || c || c || c || c |}
      \hline
    $\sqrt{s}$ (TeV)\;/\;L  (ab${\small^{-1}}\!)$         & 1.5 / 1.2  &   3  / 4.4  & 6 / 12     & 10  / 20    & 14  / 33 & 30  / 100    \\
  \hline\hline
    \multicolumn{7}{|c|}{$\sigma_{SM}$         (ab) \;\;[$N_{\rm ev}$]}\\
 \hline\hline
   $\sigma^{\rm tot}$                        & 0.03 [0] & 0.31 [1] & 1.65 [20] & 4.18 [84] & 7.02 [232] & 18.51 [1851] \\
  $\sigma(M_{HHH}\!<3$\, TeV)          & 0.03 [0] & 0.31 [1] & 1.47 [18] & 2.89 [58] & 3.98 [131] &  6.69 [669]  \\
  $\sigma(M_{HHH}\!<1$\, TeV)                           & 0.02 [0] & 0.12 [1] & 0.26 [3]  & 0.37 [7]  & 0.45 [15]  &  0.64 [64]   \\
 \hline
\end{tabular}
\caption{\label{tab:HHHnocut} Cross sections and (in squared brackets) event numbers  for triple Higgs  production via the process
$\mu^+ \mu^- \to H H H \nu \overline{\nu}$, at  collision energies and integrated
luminosities  as from table~\ref{tab:sL}. A cut $M_{\bar\nu\nu}\gtrsim 150$GeV is applied.
The effect  of imposing an upper cut on the $HHH$ invariant mass  is also detailed. Cross sections and corresponding event numbers refer to the SM case.}
\end{center}
\end{table}

A second set of relevant information is provided in table~\ref{tab:HHHnocut}, where we report the $\mu^+ \mu^- \to H H H \nu \overline{\nu}$ total cross sections and event numbers~\footnote{A cut $M_{\bar\nu\nu}\gtrsim 150$ GeV will be implicit from now on.} for the reference set of collision energies and integrated luminosities of table~\ref{tab:sL}. 
In addition to total cross sections, also the number of events close to threshold, i.e., 
with a requirement on the $HHH$-invariant-mass ($M_{HHH}$) to be less than 1 and 3 TeV is given.  
As we will discuss in the following, the sensitivity to the quartic coupling depends rather strongly on the phase space region occupied by the Higgs bosons in the final state, being the strongest close to threshold. 

Given the  very small cross section  at 1.5~TeV (cf.  table~\ref{tab:HHHnocut}), we will not
consider this option in our sensitivity studies. 
On the other hand, in section~\ref{sect:higgs4}  we will include 
the case  $\sqrt{s}=$3~TeV even if the corresponding nominal luminosity is still too low to have a reasonable event statistics.
Indeed, we will show how a (presently unrealistic) increase of a factor about 20
in luminosity (giving an integrated luminosity of   
100 ab$^{-1}$) might make  even a 
$\sqrt{s}=$3~TeV  muon collider (or, equivalently,  
a CLIC collider at 3~TeV of comparable luminosity) sensitive to a $\delta_4$ variation.

\begin{figure}
  \includegraphics[width=0.45\textwidth]{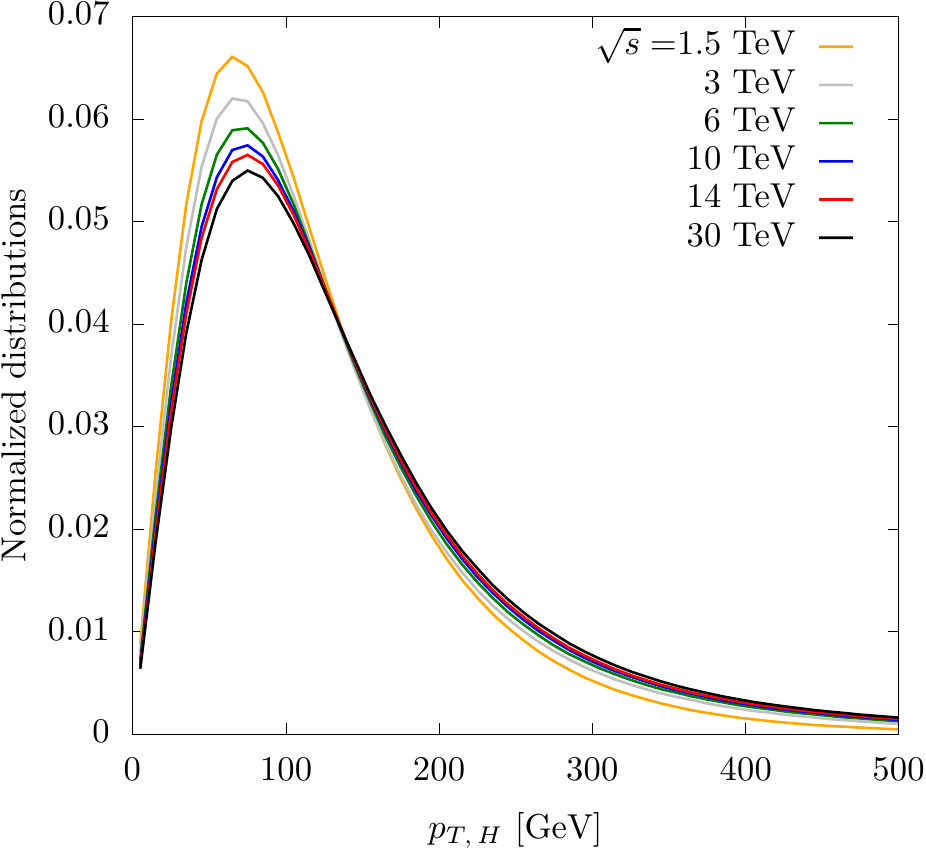}
  \hskip 0.5 cm
  \includegraphics[width=0.45\textwidth]{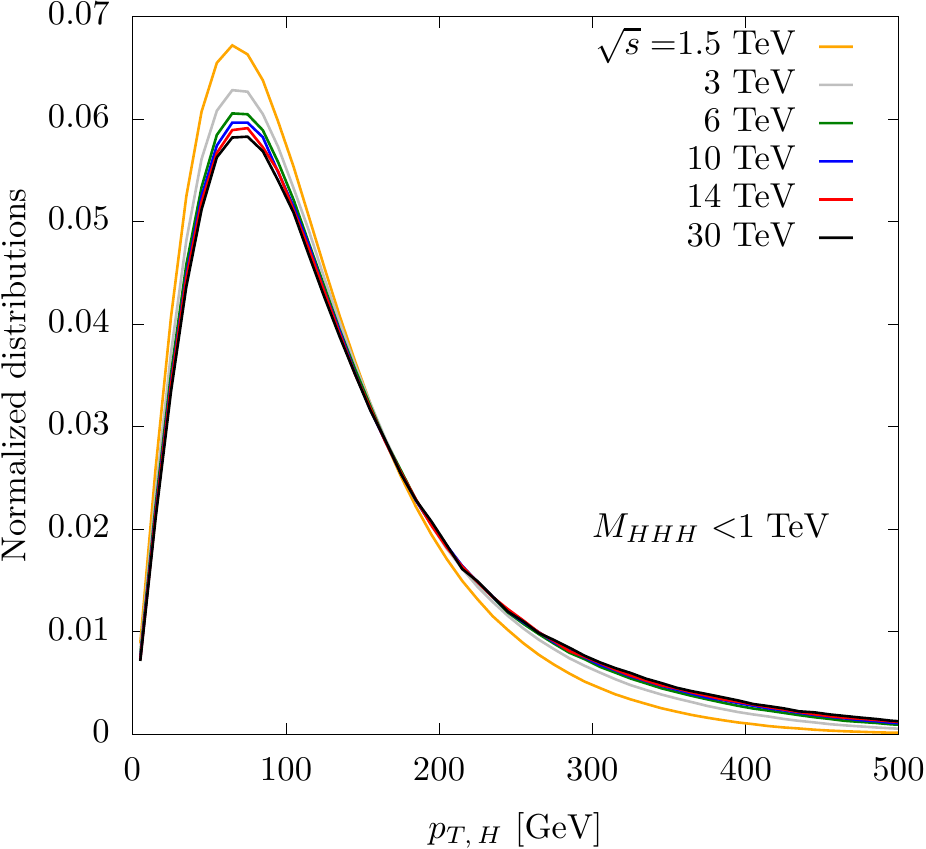}
   \vskip 0.4 cm   
  \caption{Inclusive Higgs transverse momentum distributions (normalized) for the  $\mu^+ \mu^- \to H H H \nu \overline{\nu}$ process, in the SM, at different collision energies. A technical cut of $M_{\bar\nu\nu}\gtrsim 150$ GeV is included.  The plot on the right includes an upper cut of 1 TeV on the $HHH$ invariant mass. }
  \label{fig:pth}
\end{figure}
\begin{figure}
  \includegraphics[width=0.45\textwidth]{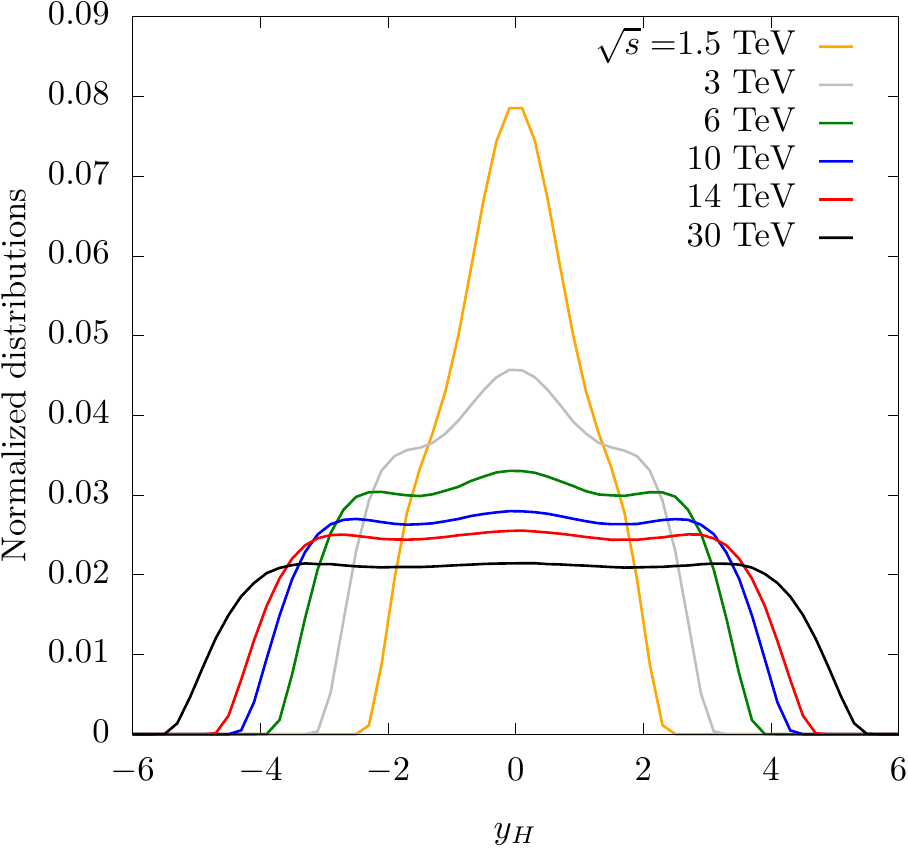}
  \hskip 0.5 cm
  \includegraphics[width=0.45\textwidth]{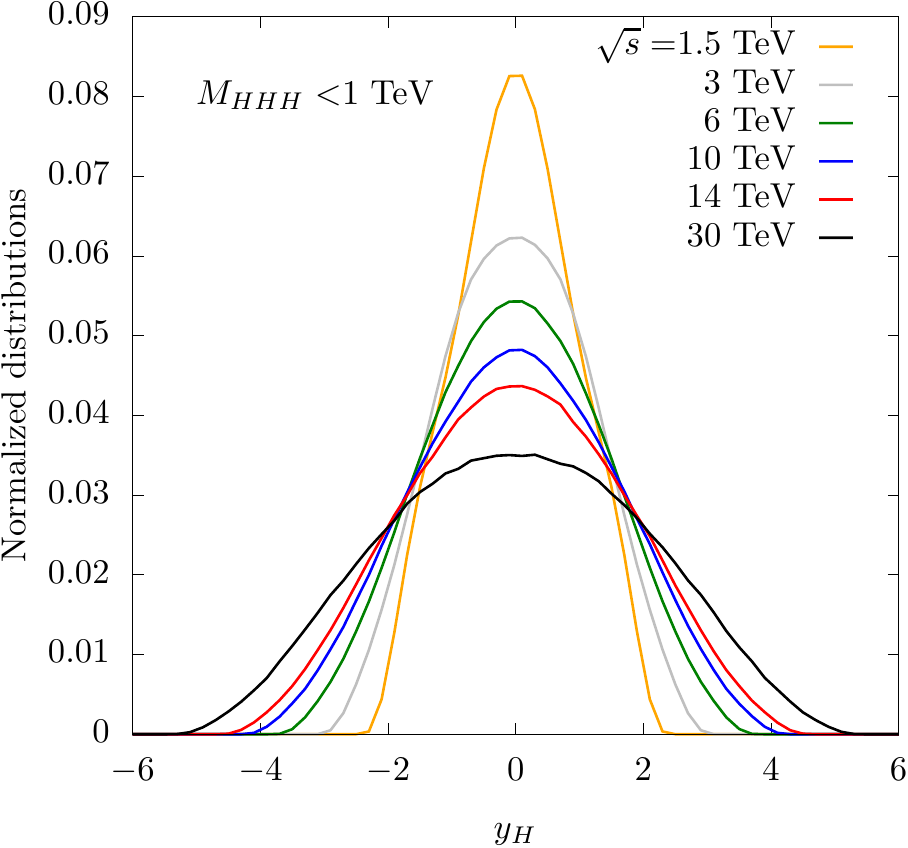}
  \caption{Inclusive Higgs rapidity distributions (normalized) for the  $\mu^+ \mu^- \to H H H \nu \overline{\nu}$ process, in the SM,  at different collision energies. A technical cut of $M_{\bar\nu\nu}\gtrsim 150 $  GeV is included. 
  The plot on the right includes an upper cut of 1 TeV on the $HHH$ invariant mass. }
  \label{fig:yh}
\end{figure}
\begin{figure}
  \includegraphics[width=0.45\textwidth]{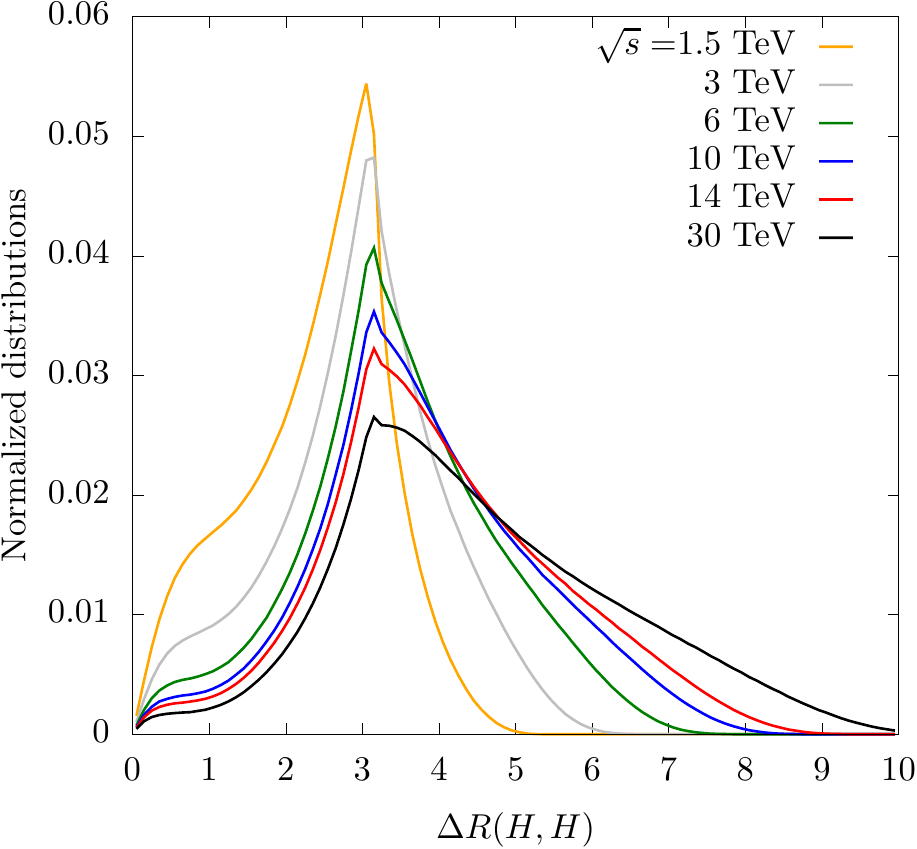}
  \hskip 0.5 cm
  \includegraphics[width=0.45\textwidth]{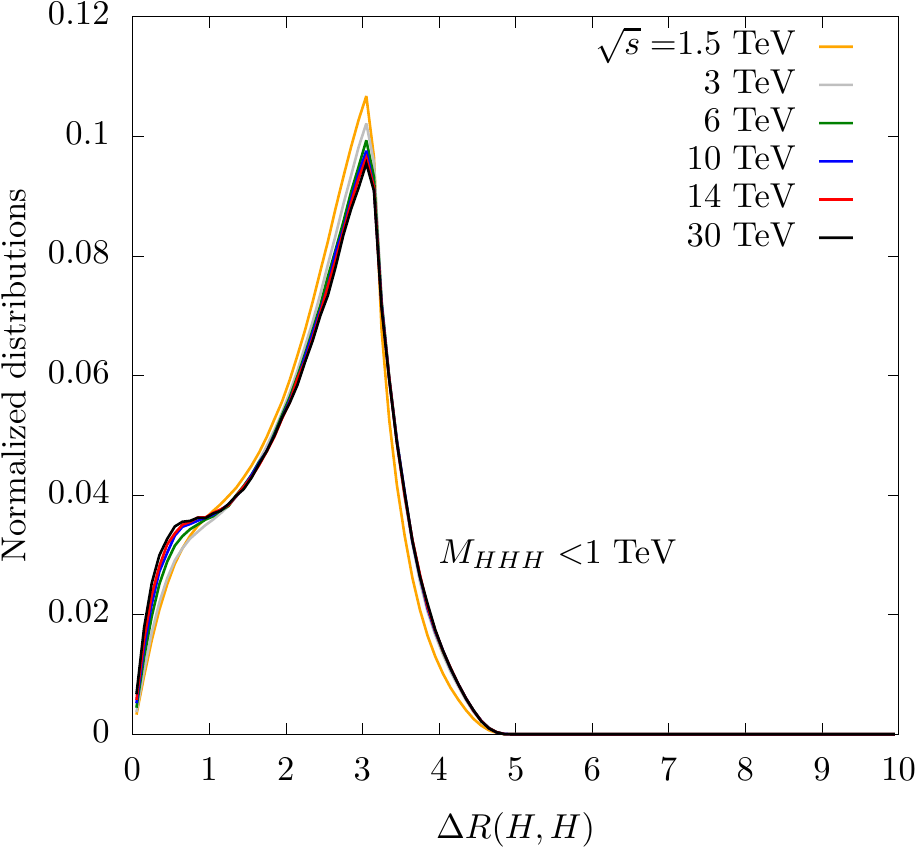}
  \caption{Inclusive $\Delta R$ distributions (normalized) for the  $\mu^+ \mu^- \to H H H \nu \overline{\nu}$ process, in the SM,  at different collision energies. A technical cut of $M_{\bar\nu\nu}\gtrsim 150$ GeV is included. The plot on the right includes an upper cut of 1 TeV on the $HHH$ invariant mass. }
    \label{fig:dr}
\end{figure}

In figures~\ref{fig:pth},~\ref{fig:yh},~\ref{fig:dr} we plot the inclusive Higgs transverse momentum, the Higgs rapidity and the Higgs-pair $\Delta R$ distributions, with and without an upper cut of 1 TeV on the $HHH$ invariant mass, respectively. 
We note that peak value of the transverse momentum is around 100 GeV, a value that turns out to be rather independent on the collider energy. The invariant mass cut at 1 TeV has a mild effect and only on the shapes of the distributions at higher energy collisions.  On the other hand, the rapidity distributions are found to have a rather strong dependence on the collision energy and also on being at threshold. At high collision energy, the rapidity range become quite large reaching more than five units in rapidity. To be efficiently detected, such  Higgs bosons would need a quite wide rapidity coverage of the detector. Finally, figure~\ref{fig:dr} shows that the most probable distance between two Higgs bosons is around $\pi$, extending to larger values at high energy, due to forward-backward Higgs production. At threshold, there is a very mild dependence on the collision energy. 

\begin{figure}
  \includegraphics[width=0.45\textwidth]{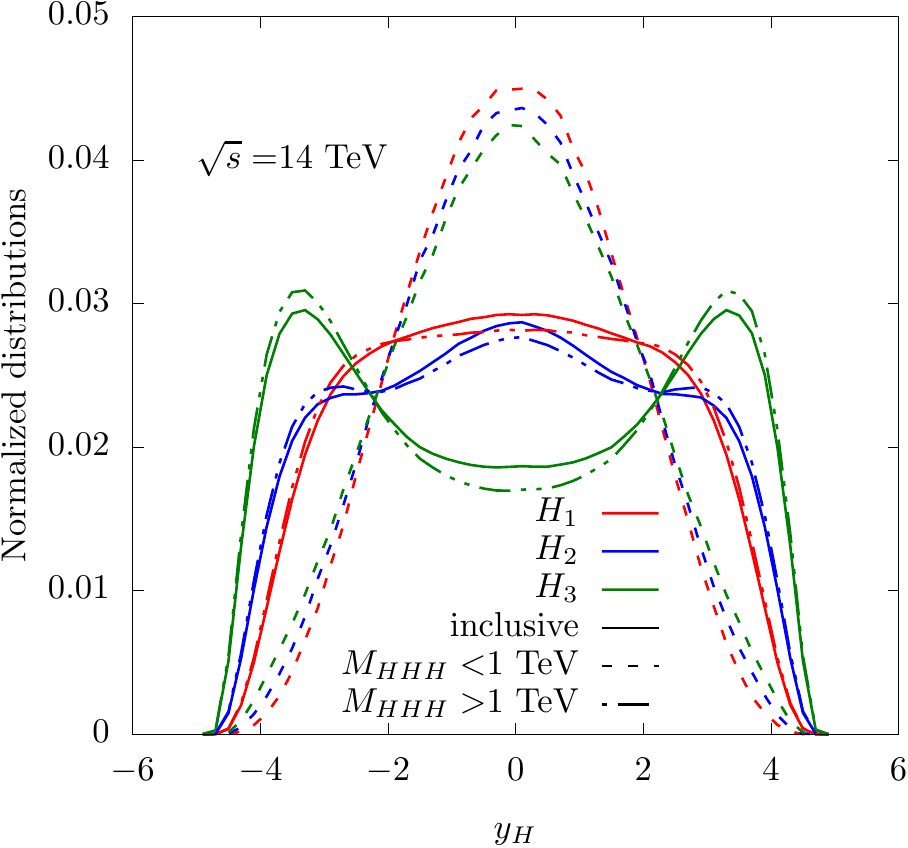}
  \hskip 0.5 cm
  \includegraphics[width=0.45\textwidth]{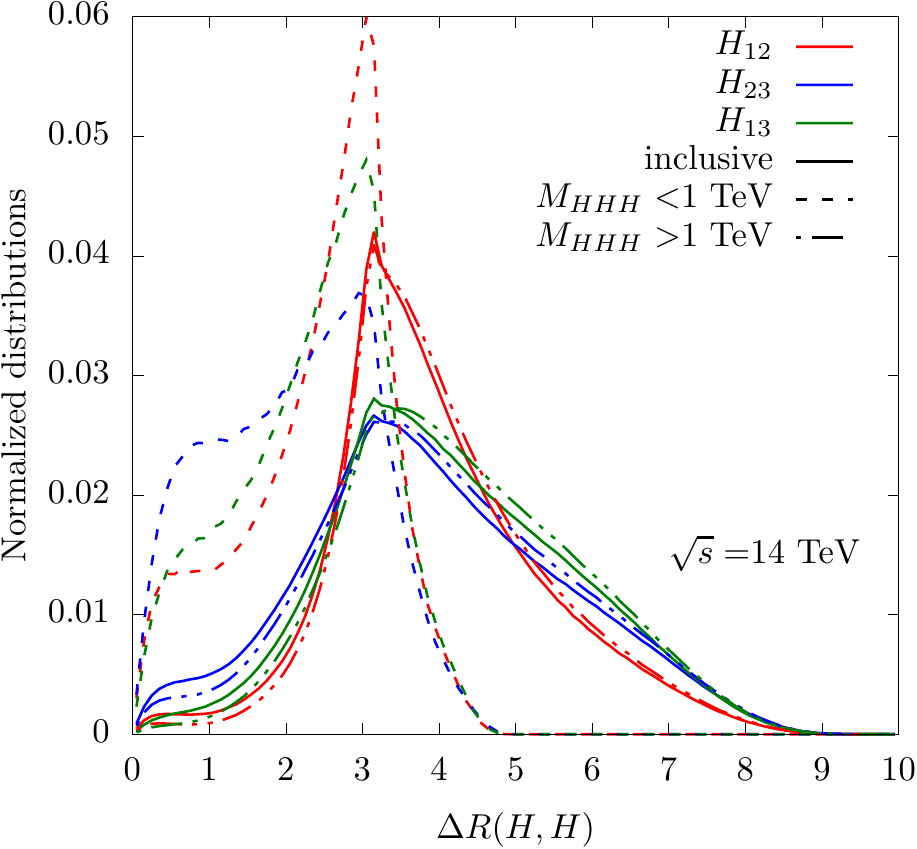}
  \caption{Higgs rapidity  (left) and Higgs-pair $\Delta R$ (right)
  distributions in $\mu^+ \mu^- \to H H H \nu \overline{\nu}$, in the SM,
  at $\sqrt s\simeq$14 TeV, for $M_{\bar\nu\nu}\gtrsim 150$ GeV. 
  The index $1$ refers to the highest-$p_T$ Higgs, while the index $3$ refers to the lowest-$p_T$ Higgs.  The solid lines stand for the inclusive distributions,
  the dashed (dotdashed) lines correspond to applying a further cut  
  \mbox{$M_{HHH}\!< (>)1$}~TeV.}
  \label{fig:dis}
\end{figure}

In order to have a more complete understanding of the dynamics of a $HHH$ event, in figure~\ref{fig:dis} we present the rapidity and $\Delta R$ distributions of each of the Higgs bosons ordered in $p_T$. The solid curves represent the inclusive sample with no lower or upper cut of 1 TeV on the $M_{HHH}$. By inspecting the two plots one concludes that at threshold the ordering of the Higgs in $p_T$ has mild effect as the Higgs have comparable momenta. On the other hand, in far from threshold configurations, which dominate inclusive cross sections, two Higgs bosons are typically rather central and back-to-back,  while the softest one is forward. 

The plots in figure~\ref{fig:yh}, obtained without imposing
any acceptance cut to the final state (apart from identifying the threshold region by the cut 
$M_{HHH}<$1~TeV) show that at high energy the Higgs bosons tend to be produced in the forward region.
At a muon collider, however, it is necessary to shield the detectors 
in the forward regions, in order to suppress
the beam-induced background. As already mentioned in the introduction, this issue was addressed for the 1.5~TeV setup, see~\cite{Foster:1995ru,Johnstone:1996hp,Mokhov:2011zzd,PhysRevSTAB.14.061001,Mokhov:2014hza}, 
where it was proposed to put nozzles with an angular opening between 5 and 10 degrees at the
points where the beam enters the detector.
Since at higher c.m. energies the beam induced radiation is expected to be more forward,
the effective angular opening of the noozles might be decreased for the largest $\sqrt{s}$
we are considering. 

In order to study the impact of these
limitations on the measurement of triple-Higgs production at the muon collider, in table~\ref{tab:acceptances1}
we collect the geometric acceptances corresponding to the energy setups in table~\ref{tab:sL}, 
for different cuts on the transverse momentum and on the rapidity of the Higgs decay products. We consider two-body Higgs decays, which give the dominant contribution to the Higgs decay width. 
In particular, in our simulation the three Higgs bosons are still produced on-shell, and then 
decayed afterwards to $b\bar{b}$ pairs, with the acceptance cuts applied to the six final-state $b$ quarks.
The scaling of the acceptances as a function of the transverse-momentum and rapidity cuts
can be read from table~\ref{tab:acceptances1}. Concerning the dependence on the rapidity cuts,
this is stronger at higher energy. For example, if one requires $p_T^b >$20~GeV,
the acceptance in the 30~TeV setup for $|\eta|<$6 is 69\% and goes down to 8\% for $|\eta|<$3,
while for $\sqrt{s}=$6~TeV it goes from 70\% to 32\%, when moving the maximum rapidity
from 6 to 3.
In the lower part of table~\ref{tab:acceptances1}, we consider the additional cut $M_{HHH}<1$~TeV.  
In this case the impact of the rapidity cut is milder, as can be expected
by looking at figure~\ref{fig:yh}. For instance, for $\vert y \vert < 3.0$
and $p_T^b > 20$~GeV the acceptance at $\sqrt{s} = 30$~TeV is about 8\%
without imposing any cut on the $HHH$ invariant mass and becomes about
37\% when the threshold region is considered. 

We stress that, in table~\ref{tab:acceptances1}, we  consider the purely geometric acceptances (i.e.
we do not apply any branching ratio), which corresponds to work under the
assumption that all the most relevant Higgs decay modes can be detected. Table~\ref{tab:acceptances1} shows
how the cuts on the transverse momentum and rapidity in the case of $6b$-jet final state reduce the expected
number of signal events, 
and point out the importance of having the largest possible rapidity coverage in the detector
in order to measure the process under consideration. This point should be carefully taken into account in future detector studies.

\begin{table}
  \begin{center}  
    \begin{tabular}{ | l | c | c | c | c | c |}
      \hline
      $\sqrt{s}$ [TeV] & $3$ TeV & $6$ TeV & $10$ TeV & $14$ TeV & $30$ TeV \\
      \hline
      \multicolumn{6}{|c|}{ $p_T^b>0$ GeV }\\ \hline
      $|y|< 2.5$ &   0.4454 [0] &   0.2123 [4]  &   0.1235 [10] &   0.0880 [20]  &   0.0431 [80]\\ \hline
      $|y|< 3.0$ &   0.7055 [1] &   0.4171 [8]  &   0.2626 [22] &   0.1932 [45]  &   0.0990 [183]\\ \hline
      $|y|< 4.0$ &   0.9542 [1] &   0.8548 [17] &   0.7086 [60] &   0.5904 [137] &   0.3589 [664]\\ \hline
      $|y|< 5.0$ &   0.9957 [1] &   0.9808 [20] &   0.9522 [80] &   0.9167 [213] &   0.7521 [1392]\\ \hline
      $|y|< 6.0$ &   0.9999 [1] &   0.9987 [20] &   0.9950 [84] &   0.9899 [230] &   0.9585 [1774]\\ \hline
      \multicolumn{6}{|c|}{ $p_T^b>20$ GeV }\\  \hline
      $|y|< 2.5$ &   0.3431 [0] &   0.1636 [3]  &   0.0956 [8]  &   0.0685 [16]  &   0.0338 [63]\\ \hline
      $|y|< 3.0$ &   0.5355 [1] &   0.3178 [6]  &   0.2004 [17] &   0.1479 [34]  &   0.0763 [141]\\ \hline
      $|y|< 4.0$ &   0.6925 [1] &   0.6335 [13] &   0.5292 [44] &   0.4419 [103] &   0.2695 [499]\\ \hline
      $|y|< 5.0$ &   0.7013 [1] &   0.7003 [14] &   0.6885 [58] &   0.6681 [155] &   0.5542 [1026]\\ \hline
      $|y|< 6.0$ &   0.7013 [1] &   0.7018 [14] &   0.7018 [59] &   0.7019 [163] &   0.6896 [1276]\\ \hline
      \multicolumn{6}{|c|}{ $p_T^b>0$ GeV, $M_{HHH}<1$ TeV }\\ \hline
      $|y|< 2.5$ &   0.6712 [0] &   0.5396 [1] &   0.4563 [1] &   0.4118 [2]  &   0.3309 [7]\\ \hline
      $|y|< 3.0$ &   0.8572 [1] &   0.7479 [1] &   0.6557 [2] &   0.5984 [4]  &   0.4889 [14]\\ \hline
      $|y|< 4.0$ &   0.9821 [1] &   0.9573 [3] &   0.9182 [6] &   0.8794 [11] &   0.7697 [38]\\ \hline
      $|y|< 5.0$ &   0.9988 [1] &   0.9955 [3] &   0.9893 [7] &   0.9816 [14] &   0.9420 [56]\\ \hline
      $|y|< 6.0$ &   1.0000 [1] &   0.9998 [3] &   0.9992 [7] &   0.9982 [15] &   0.9924 [63]\\ \hline
      \multicolumn{6}{|c|}{ $p_T^b>20$ GeV, $M_{HHH}<1$ TeV }\\  \hline
      $|y|< 2.5$ &   0.5144 [0] &   0.4135 [1] &   0.3494 [1] &   0.3162 [2]  &   0.2537 [5]\\ \hline
      $|y|< 3.0$ &   0.6377 [1] &   0.5590 [1] &   0.4905 [2] &   0.4490 [3]  &   0.3667 [11]\\ \hline
      $|y|< 4.0$ &   0.6979 [1] &   0.6871 [2] &   0.6630 [4] &   0.6385 [8]  &   0.5610 [28]\\ \hline
      $|y|< 5.0$ &   0.6996 [1] &   0.7006 [2] &   0.6991 [5] &   0.6971 [10] &   0.6744 [41]\\ \hline
      $|y|< 6.0$ &   0.6996 [1] &   0.7007 [2] &   0.7008 [5] &   0.7018 [11] &   0.7011 [45]\\ \hline

    \end{tabular}
    \caption{\label{tab:acceptances1} Ratio of the expected number of events with and without imposing acceptance
      cuts on the Higgs decay products for the energy setups in table~\ref{tab:sL}.
      The Higgs bosons are produced on-shell and decayed to $b\bar{b}$ pairs but no branching ratio is applied.
      The numbers in square parenthesis are the expected number of events for the process
      $\mu^+\mu^- \to HHH\nu\bar{\nu}$ (the $H\to b\bar{b}$ branching ratio is not included) 
      after imposing the cuts when assuming the integrated luminosities in table~\ref{tab:sL} for the c.m.
      energies under consideration.  See main text for details.}
  \end{center}
\end{table}

\section{Triple Higgs production with anomalous self-couplings }

We can now pass to consider in detail how modifications of the trilinear and quartic couplings can modify cross sections and distributions. 
As already mentioned, the Feynman diagrams contributing to the process
$\mu^+ \mu^- \to H H H \nu \overline{\nu}$ can involve one quartic Higgs vertex or
up to two Higgs trilinear vertices, see figure~\ref{Feyndiags}.

As a result, the most general expression for the cross section as a function of the deviations from the SM cubic and quartic Higgs couplings can be expressed in terms of a polynomial which is quartic in $\delta_3$ and quadratic in $\delta_4$:
\begin{equation}
     \sigma= c_1
     +       c_2 \delta_3 
     +       c_3 \delta_4
     +       c_4 \delta_3  \delta_4
     +       c_5 \delta_3^2
     +       c_6 \delta_4^2
     +       c_7 \delta_3^3
     +       c_8 \delta_3^2 \delta_4
     +       c_9 \delta_3^4 \,,
\label{eq:sigmak3k4}
\end{equation}
where the coefficients $c_i$ can be obtained once for all from a MC simulation and they are collected in table~\ref{tab:ci}, for the total cross sections with and without an upper cut on the $HHH$ invariant mass of 1 TeV. This parametrization is useful for at least two reasons. The first is that it can be used to extract sensitivities to different scenarios  without the need to rerun MC simulations for each benchmark point. The second advantage is that it is possible to directly gauge the sensitivity to new physics effects by comparing the value of the SM coefficient ($c_1$), with the linear terms $c_2,c_3$, which are dominant for $\delta_{3,4} \ll 1$, and  the quadratic (mixed or diagonal) terms ($c_{4,5,6}$), the cubic ($c_{7,8}$) and finally the quartic terms ($c_9$). First, the SM coefficient, as we had already seen in figure~\ref{total}, grows faster than linearly, yet tends to flatten at high energy. As also seen before, the increase of the cross section is clearly provided by configurations which are far from threshold, and where at least one Higgs boson is soft and can be very forward. In fact, once an upper cut on the $HHH$ invariant mass of 1 TeV is set, the increase on the cross sections is less than linear and very mild.   Second, at the linear level and for total cross sections, the sensitivity to $\delta_4$ is smaller than the one to $\delta_3$ by a factor ranging between 2 to 100. On the other hand, if one focuses on events at threshold, there is a rather uniform difference of only a factor of two, the sign being opposite. This generically implies that positively correlated changes of the  $\delta_4$ and $\delta_3$, will be more difficult to constrain than variations in opposite directions. For example,  in the SMEFT case where $\delta_4=6\, \delta_3$,  there will be a cancellation, yet with the $\delta_4$ contribution being dominating. 
More in general,  the difference between the sensitivity at the inclusive level and at threshold,  entails the possibility for flat directions in the parameter space to be lifted.  Third, in presence of larger deviations, the higher-order terms in the polynomial could become the dominant effects. 
In this case, one notices that $c_6$, corresponding to the $\delta_4^2$ term, is always smaller than $c_4$, the coefficient of the $\delta_3 \delta_4$ term. This means that a joint departure of the trilinear and quartic term will be in general easier to detect, than that of the quartic alone. 

\begin{table}
\begin{center}
  \begin{tabular}{| c | c | c | c | c | c | c | c | c | c |}
    \hline
      \hline
    \multicolumn{6}{|c|}{$   \sigma= c_1
     +       c_2 \delta_3 
     +       c_3 \delta_4
     +       c_4 \delta_3  \delta_4
     +       c_5 \delta_3^2
     +       c_6 \delta_4^2
     +       c_7 \delta_3^3
     +       c_8 \delta_3^2 \delta_4
     +       c_9 \delta_3^4$}\\
            \multicolumn{6}{|c|}{ }\\
\hline\hline
    $\sqrt{s}$\;(TeV) & 3 & 6 & 10 & 14 & 30 \\
   \hline \hline
  \multicolumn{6}{|c|}{$c_i\;\;$   (ab) }\\
    \hline
    $c_1$      &    0.3127 &   1.6477 &   4.1820 &   7.0200 &  18.5124  \\
\hline
    $c_2$      &   -0.1533 &  -1.7261 &  -4.4566 &  -7.1000 & -15.9445  \\
    $c_3$      &   -0.0753 &  -0.1159 &  -0.1166 &  -0.1147 &  -0.1117  \\
 \hline
    $c_4$      &   -2.0566 &  -6.3052 & -11.4981 & -15.9807 & -29.2794  \\
    $c_5$      &    4.7950 &  14.9060 &  27.1081 &  37.4658 &  67.7539  \\
    $c_6$      &    0.2772 &   0.8637 &   1.5992 &   2.2455 &   4.2038  \\
\hline
    $c_7$      &   -1.8353 &  -4.3210 &  -6.6091 &  -8.3962 & -13.0964  \\
    $c_8$      &    0.5032 &   1.1861 &   1.8173 &   2.2967 &   3.5217  \\
\hline
    $c_9$      &    0.2943 &   0.5954 &   0.8946 &   1.1611 &   1.9349  \\
   \hline   \hline
  \multicolumn{6}{|c|}{$\bar c_i\equiv c_i(M_{HHH}<1$~TeV)\;\;   (ab) }\\
    \hline
    $\bar c_1$     &  0.1165 &   0.2567 &   0.3743 &   0.4541 &   0.6404 \\
\hline
    $\bar c_2$     &  0.1667 &   0.3003 &   0.4046 &   0.3545 &   0.6972 \\
    $\bar c_3$     & -0.0768 &  -0.1510 &  -0.2105 &  -0.2285 &  -0.3519 \\
 \hline
    $\bar c_4$     & -1.3604 &  -2.8996 &  -4.1522 &  -5.0582 &  -6.9538 \\
    $\bar c_5$     &  3.1017 &   6.6033 &   9.4721 &  11.4547 &  15.9505 \\
    $\bar c_6$     &  0.1842 &   0.3954 &   0.5679 &   0.6931 &   0.9543 \\
\hline
    $\bar c_7$     & -1.5210 &  -3.0591 &  -4.3186 &  -4.8598 &  -7.3196 \\
    $\bar c_8$     &  0.4222 &   0.8550 &   1.2103 &   1.3906 &   2.0398 \\
 \hline
    $\bar c_9$     &  0.2691 &   0.5482 &   0.7720 &   0.9702 &   1.2482 \\
    \hline
  \end{tabular}
  \caption{\label{tab:HHHcoeff2} 
  Coefficients $c_i$, ruling the 
 $\mu^+ \mu^- \to H H H \nu_\mu \overline{\nu}_\mu$ cross-section 
 dependence on the Higgs anomalous
  self-couplings $\delta_3$ and $\delta_4$ (as defined in the first row of the table),
 at different c.m. energies. The coefficients $\bar c_i$, entering the residual cross sections after applying a 1-TeV upper cut on the $HHH$ invariant mass,  are also detailed.  
  }
  \label{tab:ci}
\end{center}
\end{table}

\begin{figure} 
  \includegraphics[width=0.45\textwidth]{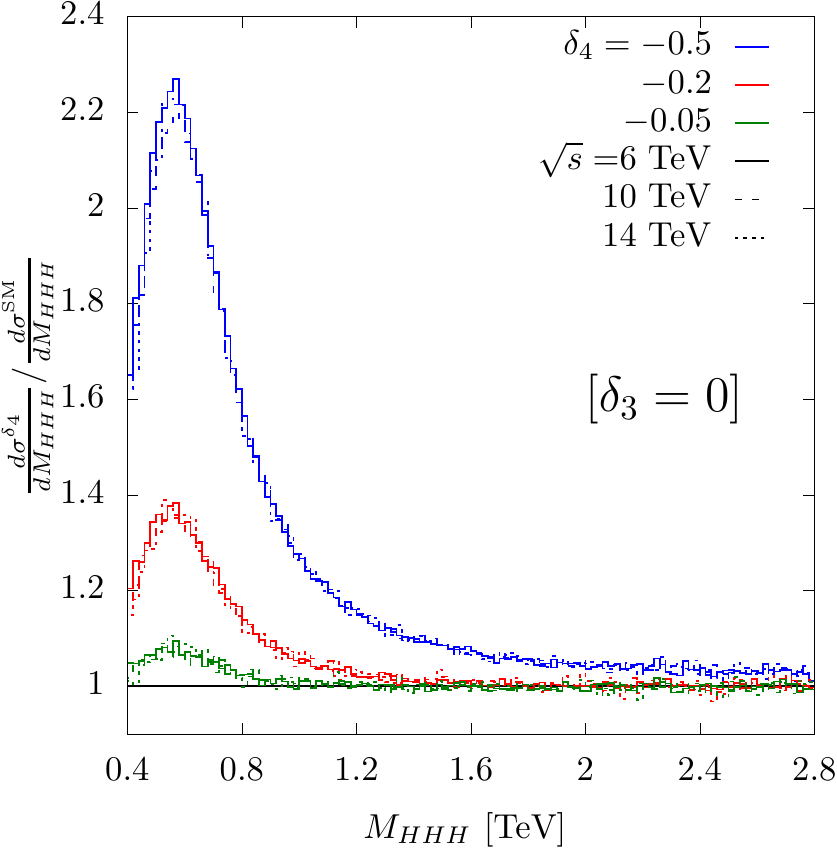}
   \hskip 0.4 cm
  \includegraphics[width=0.45\textwidth]{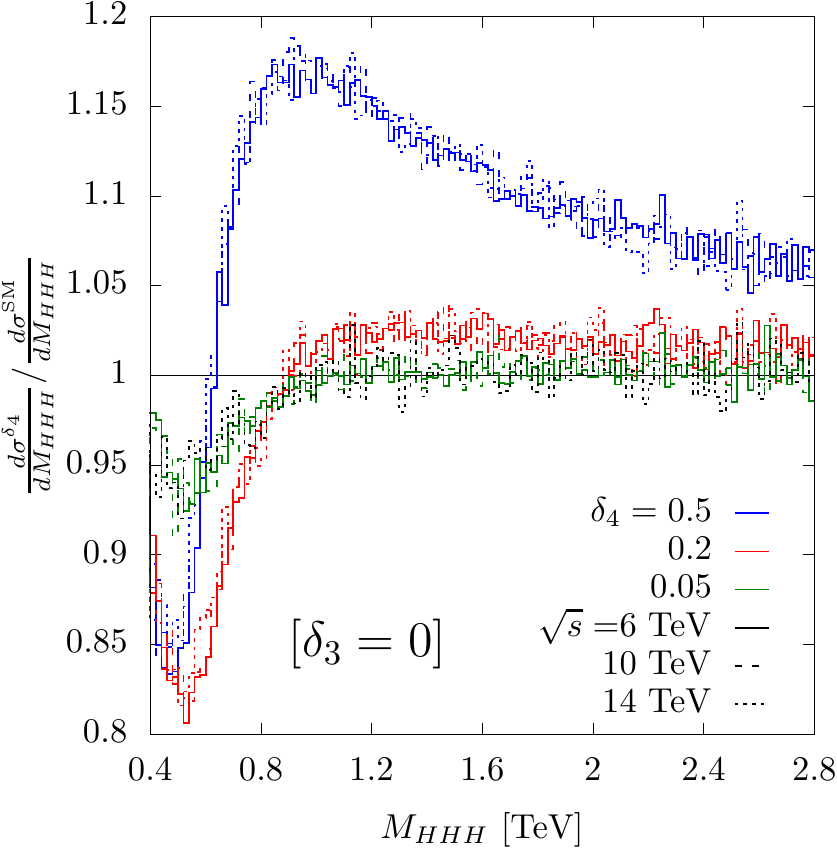}
  \caption{Dependence of the $M_{HHH}$ distributions on a variation of the quartic Higgs
  coupling,  for  three energy setups, assuming $\delta_3=0$ (i.e., a SM trilinear self-coupling).}
  \label{fig:m3h1}
  
\end{figure}

\begin{figure}
  \includegraphics[width=0.45\textwidth]{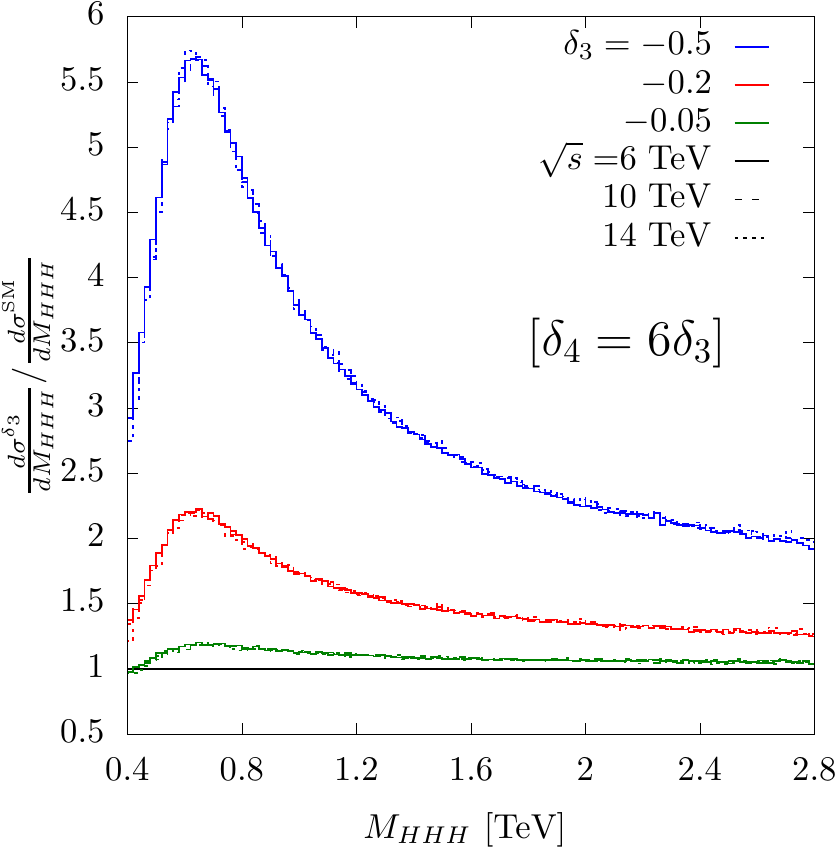}
   \hskip 0.4 cm
  \includegraphics[width=0.45\textwidth]{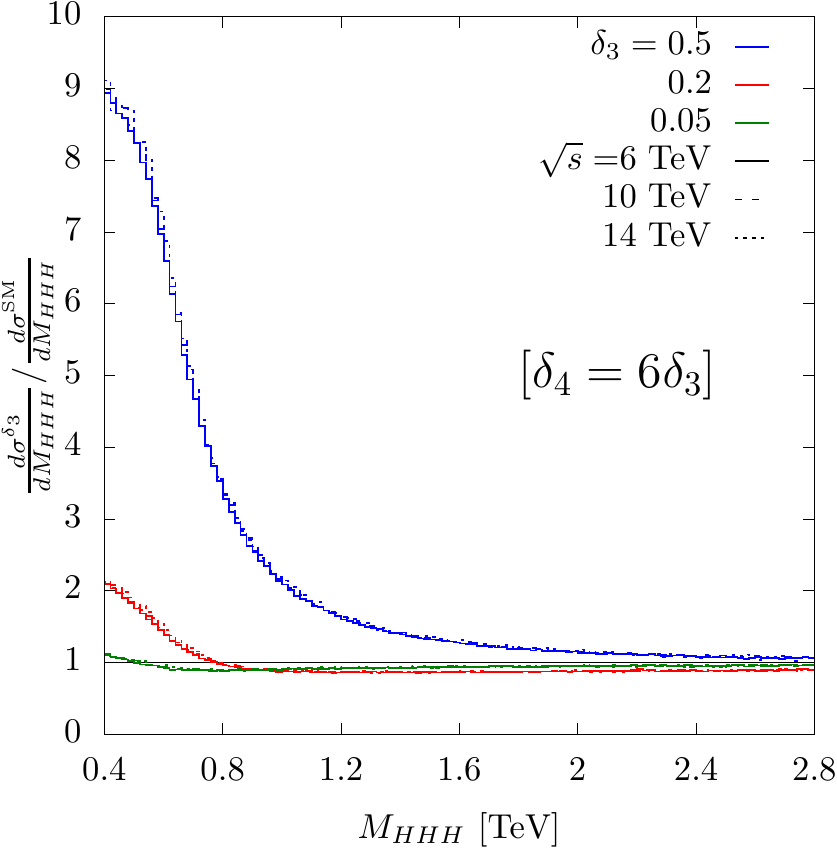}
  \caption{Dependence of the $M_{HHH}$ distributions on a variation of the trilinear Higgs coupling,
    for  three energy setups, assuming $\delta_4=6\,\delta_3$.}
  \label{fig:m3h2}
\end{figure}

Finally, we investigate the discriminating power of differential distributions, focusing our attention on the $HHH$ invariant mass. In figure~\ref{fig:m3h1} we plot the ratio between the $M_{HHH}$ distribution in a scenario where $\delta_3=0$, for 
$\delta_4 =-0.5, -0.2,-0.05$ (left plot) and for  $\delta_4 =0.5, 0.2,0.05$ (right plot) for different c.m. energies. The first observation is the size as well as the dependence of the corrections on the $M_{HHH}$ are very different between positive and negative values of $\delta_4$. The main reason can be traced back to the fact that even at the total integrated level the linear coefficient $c_3$ is negative while the quadratic coefficient $c_6$ is positive. For negative values of $\delta_4$ the contributions sum and the final result is always larger than the SM, the larger effects being at threshold. 
For positive values of  $\delta_4$, cancellations take place between the differential version of $c_3$ and $c_6$, 
leading to a final non trivial pattern shown on the right plot: corrections start negative very close to threshold, and then become positive above about 600-800 GeV. In figure~\ref{fig:m3h2} we show the results of an analogous study, assuming $\delta_3 =-0.5, -0.2,-0.05$ (left plot) and  $\delta_3 =0.5, 0.2,0.05$  with $\delta_4=6\,\delta_3$, i.e., in the SMEFT scenario. Also in this case the shape changes are larger at threshold and deviations with respect to SM predictions can be quite significant.



\section{Sensitivity to the Higgs self-coupling deviations}
\label{sect:higgs4}
We are now ready to perform the first exploration of the sensitivity of a future muon collider to deviations of the Higgs self-couplings. 

For the sake of simplicity, we restrict the presentation to two possibly relevant scenarios: 
\begin{itemize}
\item[A)] $\delta_3=0$, $\delta_4 \neq 0$, i.e., deviations only in the quartic Higgs coupling; 
\item[B)] $\delta_4=6 \,\delta_3$, i.e., the pattern of deviations as expected from the SMEFT at dim=6. 
\end{itemize}
Scenario A assumes that no deviations on the trilinear coupling have been detected (and/or exist) and explores the possibility that new physics effects appear for the first time in the quartic self-coupling. Scenario B, on the other hand, assumes the SMEFT scaling between the two couplings. This scenario would fit the case where a deviation  in the trilinear coupling is observed in other observables, such as in $HH$ production. In this situation, an interesting question would be whether the deviation in $\delta_4$ would follow the linear SMEFT pattern or not.

To provide a first estimation of the sensitivity, we focus on the signal process $\mu^+ \mu^- \to H H H \nu \overline{\nu}$ and disregard possible backgrounds. In so doing, we are clearly setting an optimal target for more detailed future phenomenological and experimental investigations. 
We define the sensitivity to the non-SM Higgs couplings as:
\begin{equation}
  \frac{|N-N_{\rm SM}|}{\sqrt{N_{\rm SM}}},
  \label{eq:significance}
\end{equation}
where $N_{\rm SM}$ is the number of events assuming  
$\delta_3=\delta_4=0$, 
while $N$ is the
number of events obtained for the values of 
$\delta_3$ and $\delta_4$ 
under consideration.
\begin{figure}
  \includegraphics[width=0.45\textwidth]{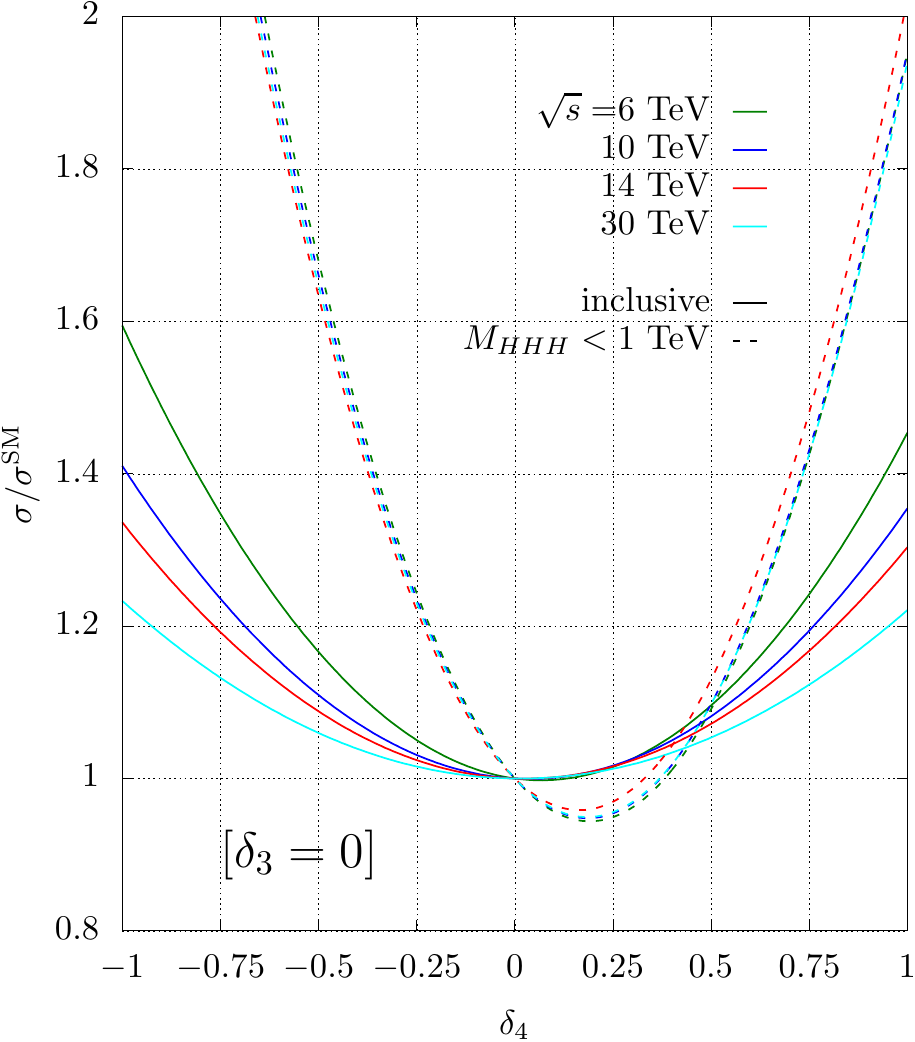}
   \hskip 0.5 cm
  \includegraphics[width=0.45\textwidth]{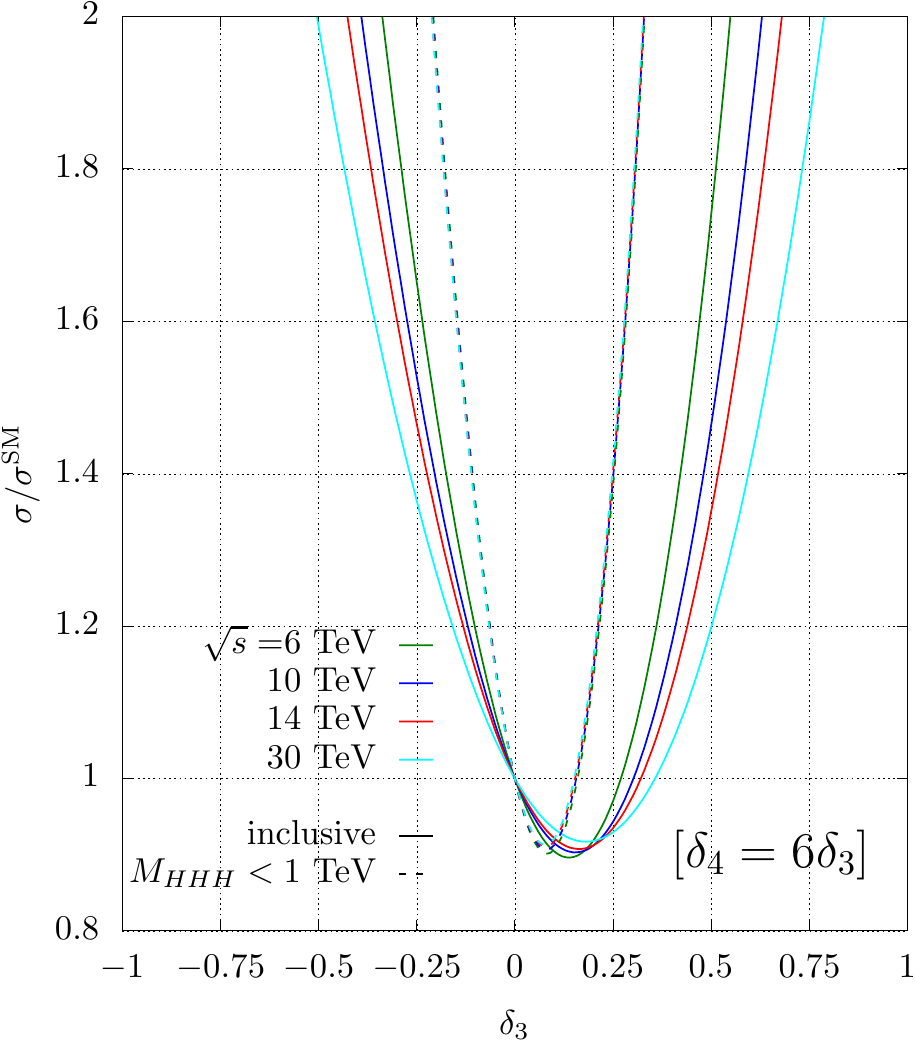}
  \caption{Dependence of the $\mu^+ \mu^- \to H H H \nu \overline{\nu}$ cross section on the 
    anomalous Higgs  self-couplings in two different scenarios: A ($\delta_3=0$) on the left and
    B ($\delta_4=6\,\delta_3$) on the right. In the latter case the ratio of the cross
    sections is expressed in terms of  $\delta_3$.}
   \label{fig:2s}
\end{figure}

\begin{figure}
  \includegraphics[width=0.45\textwidth]{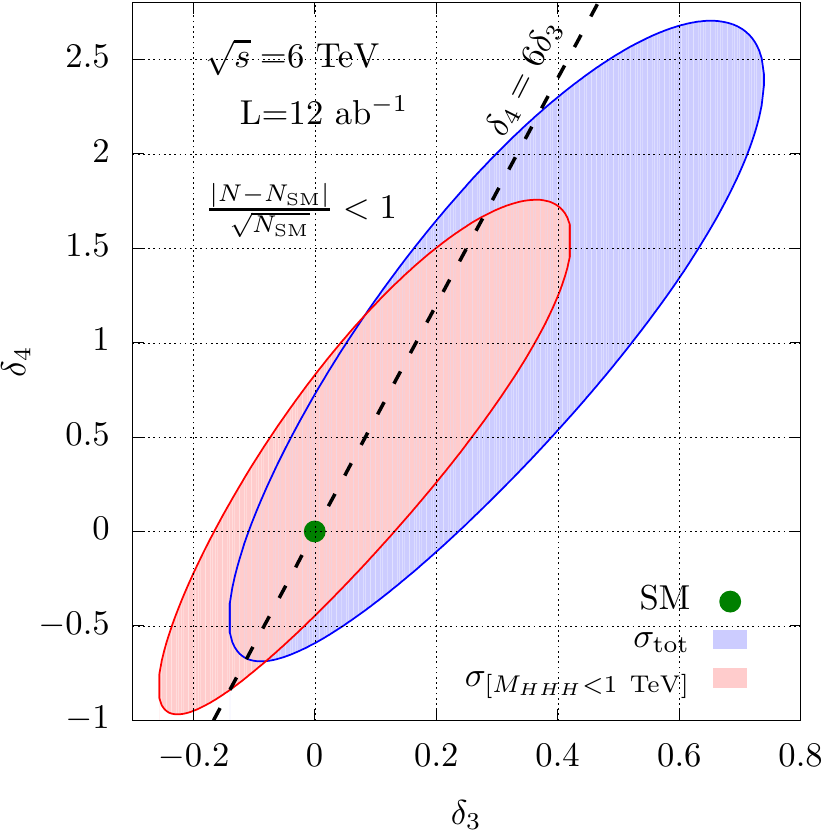}
   \hskip 0.5 cm
  \includegraphics[width=0.45\textwidth]{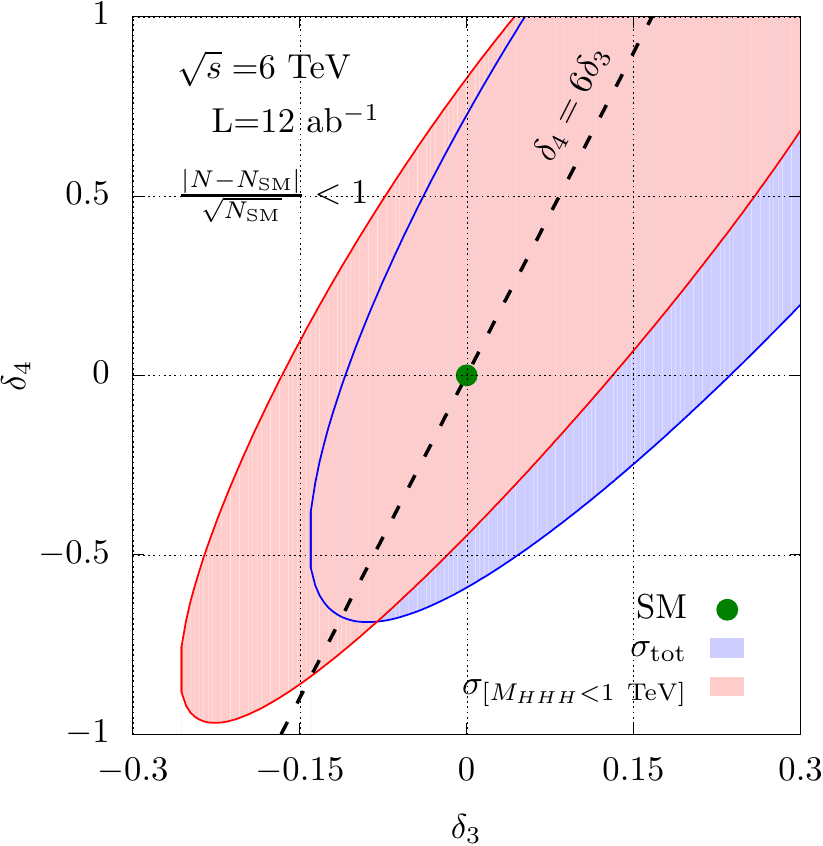}
   \vskip 0.5 cm
  \includegraphics[width=0.45\textwidth]{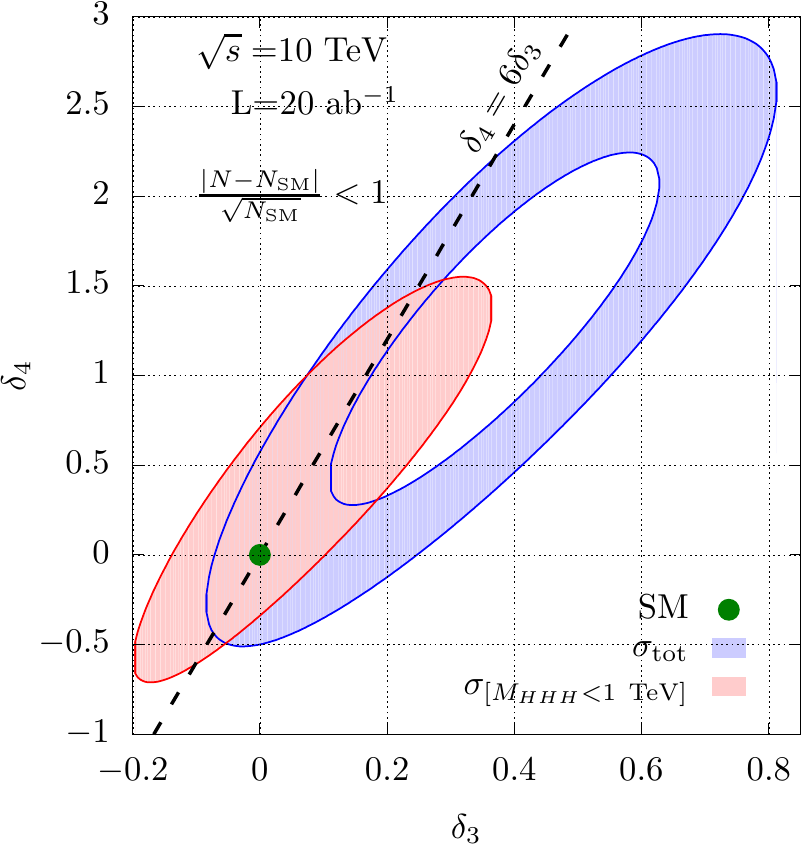}
  \hskip 0.5 cm
  \includegraphics[width=0.45\textwidth]{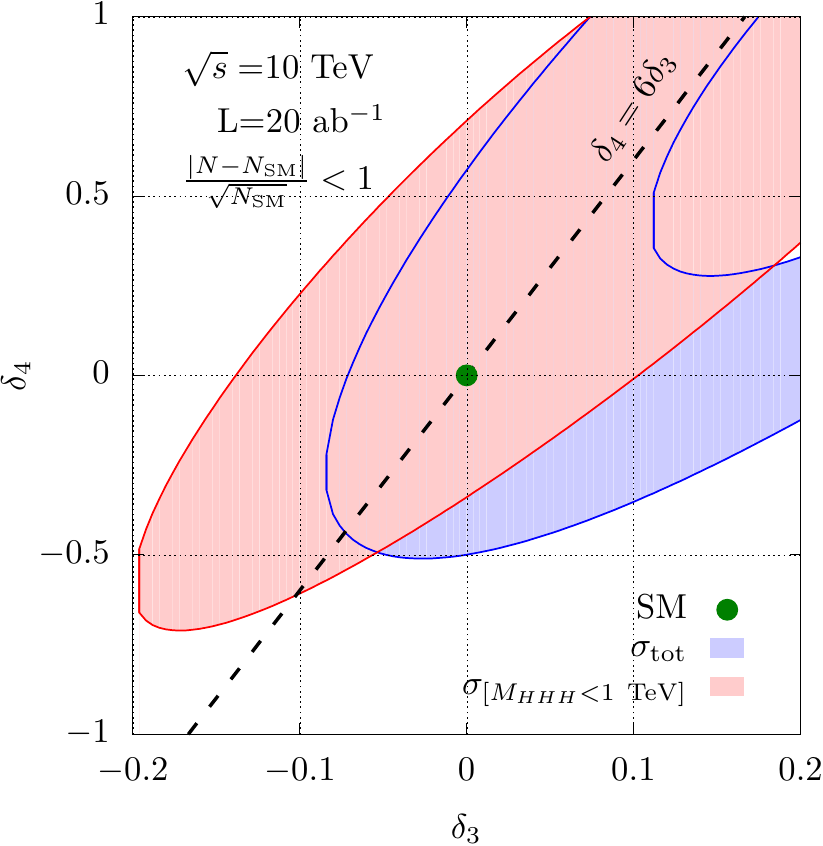}
  \caption{Left: 1-$\sigma$ exclusion plots for the anomalous Higgs self-couplings
  in terms of the standard deviations 
  $|N-N_{\rm SM}|/\sqrt{N_{\rm SM}}$
  from the SM  (green dot), where the event numbers $N$ 
  refer either to $\sigma(\mu^+ \mu^- \to H H H \nu \overline{\nu})$,  for $M_{\bar\nu\nu}\gtrsim 150$GeV (blue area), or to the same cross section
  with an upper cut of 1 TeV on the $HHH$ invariant mass (red area).
  Right: same plots zoomed around the SM configuration. 
 }
 \label{fig:lim1}
\end{figure}
\begin{figure}
   \includegraphics[width=0.45\textwidth]{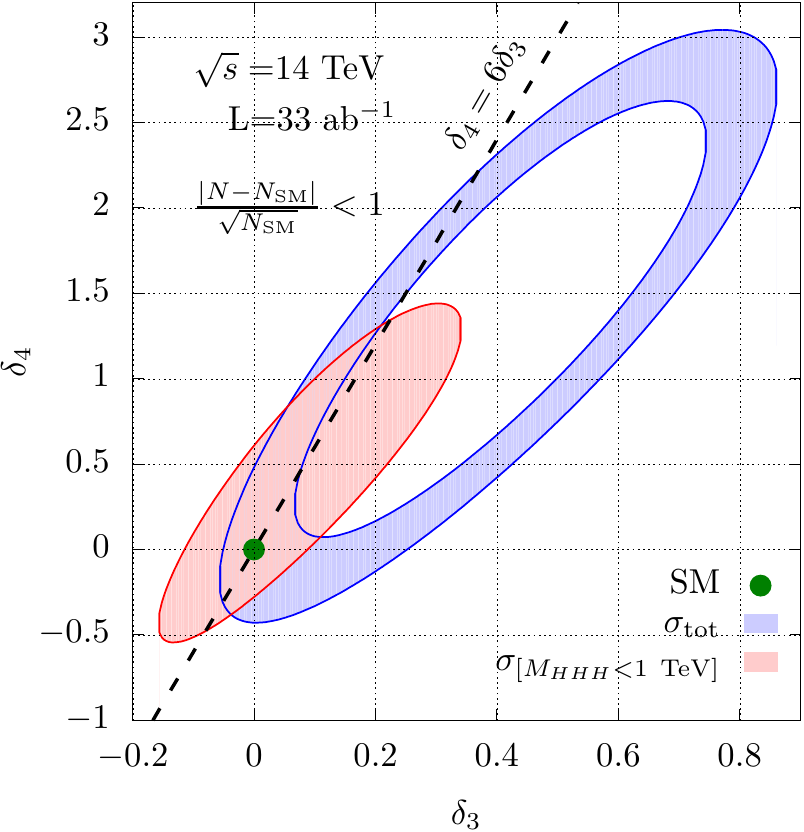}
    \hskip 0.5 cm
  \includegraphics[width=0.45\textwidth]{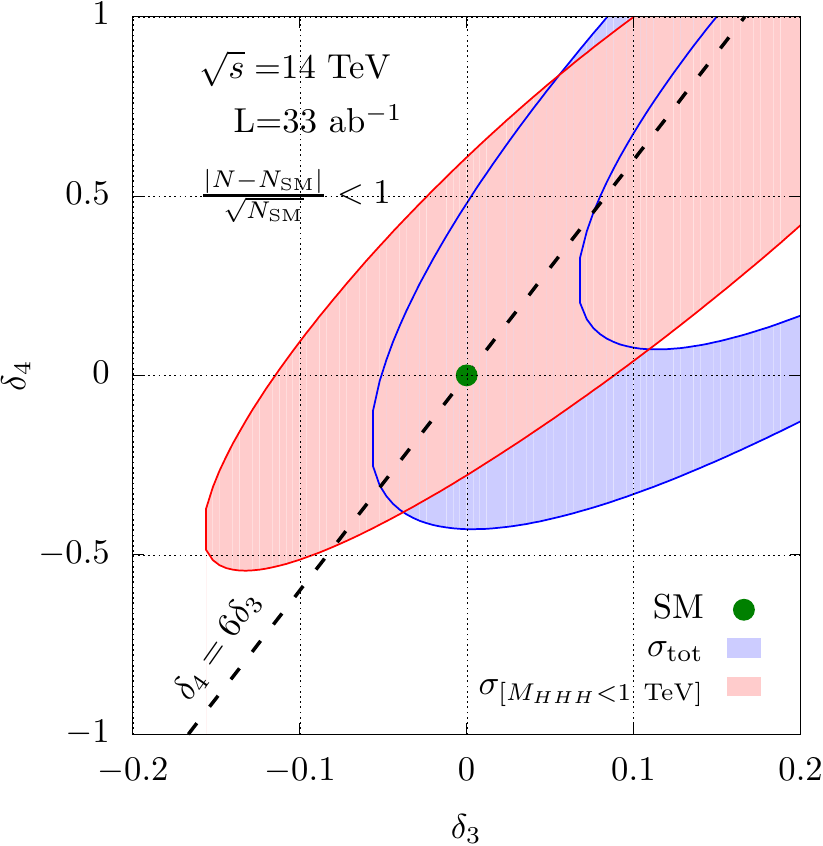}
   \vskip 0.5 cm
   \includegraphics[width=0.45\textwidth]{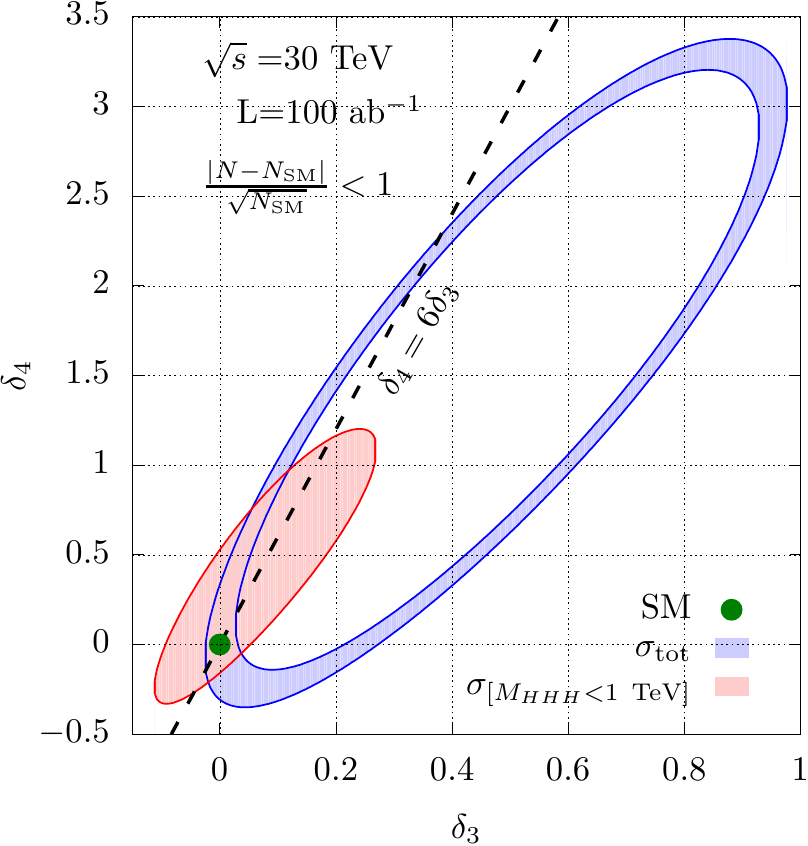}
    \hskip 0.5 cm
  \includegraphics[width=0.45\textwidth]{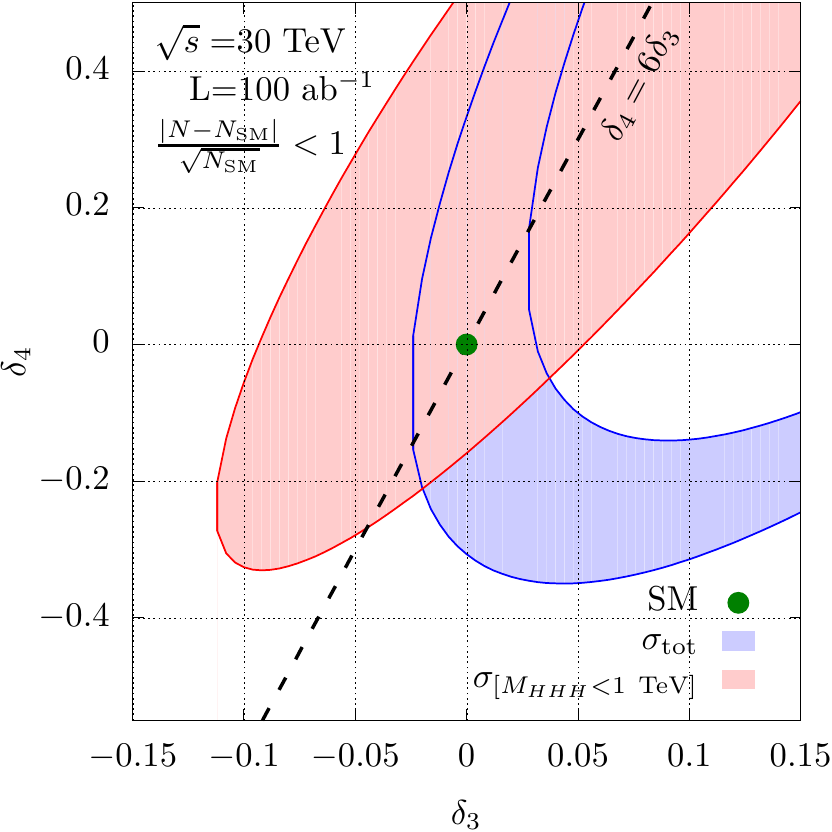}
  \caption{Left: 1-$\sigma$ exclusion plots for the anomalous Higgs self-couplings
  in terms of the standard deviations 
  $|N-N_{\rm SM}|/\sqrt{N_{\rm SM}}$
  from the SM  (green dot), where the event numbers $N$ 
  refer either to $\sigma(\mu^+ \mu^- \to H H H \nu \overline{\nu})$,  for $M_{\bar\nu\nu}\gtrsim 150$GeV (blue area), or to the same cross section
  with an upper cut of 1 TeV on the $HHH$ invariant mass (red area).
  Right: same plots zoomed around the SM configuration. 
 }
  \label{fig:lim2}
\end{figure}
\begin{figure}
  \includegraphics[width=0.45\textwidth]{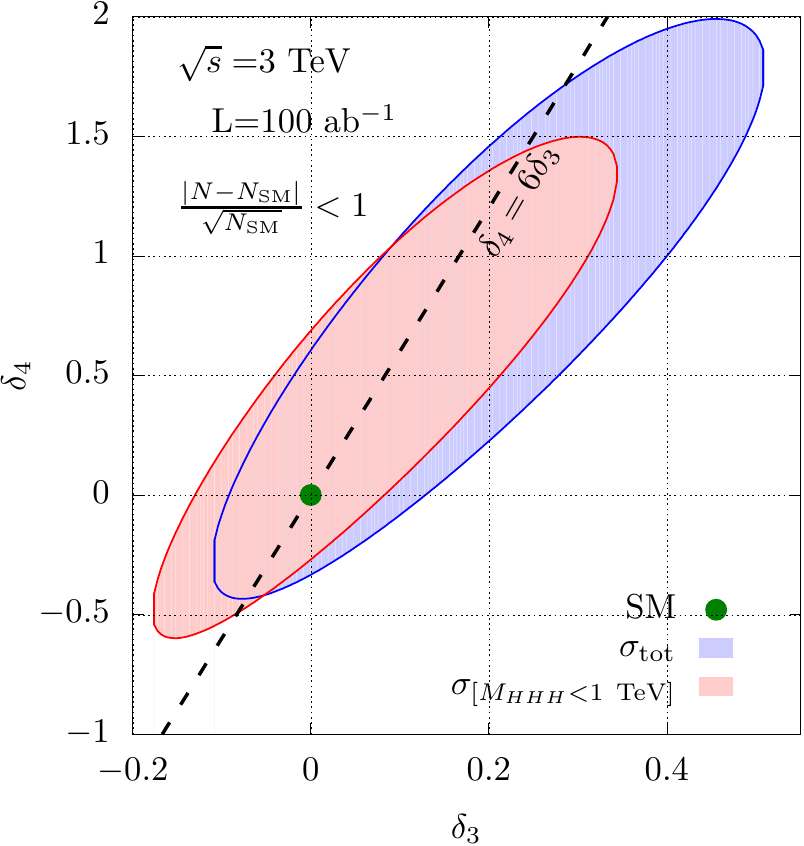}
   \hskip 0.5 cm
  \includegraphics[width=0.45\textwidth]{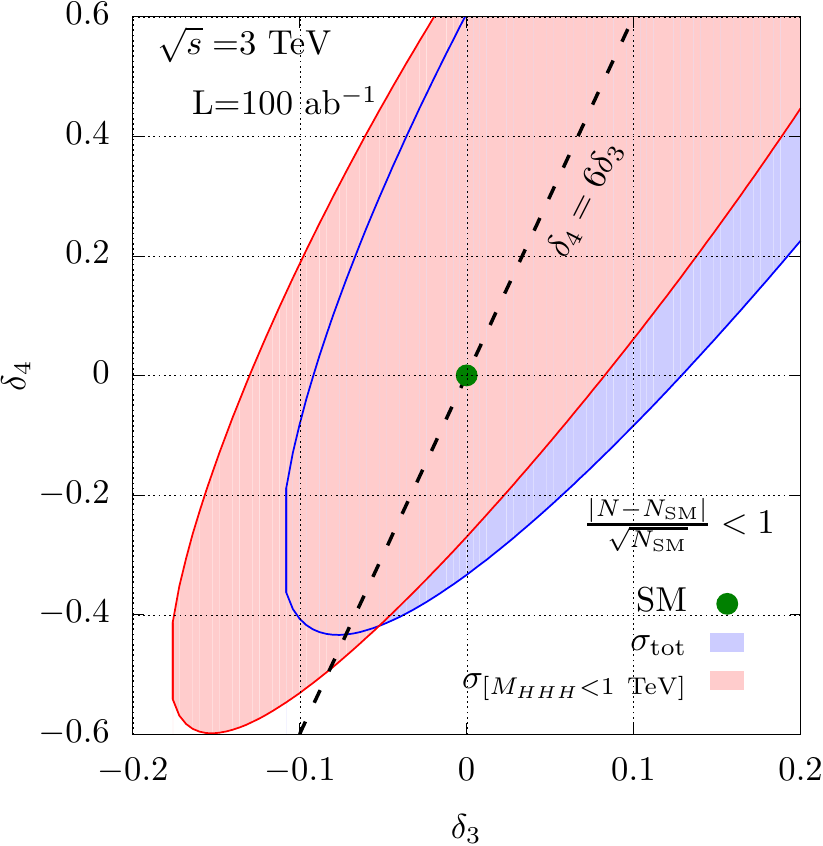}
   \caption{Left: 1-$\sigma$ exclusion plot for the anomalous Higgs self-couplings
  in terms of the standard deviations 
  $|N-N_{\rm SM}|/\sqrt{N_{\rm SM}}$
  from the SM  (green dot), where the event numbers $N$ 
  refer either to $\sigma(\mu^+ \mu^- \to H H H \nu \overline{\nu})$,  for $M_{\bar\nu\nu}\gtrsim 150$GeV (blue area), or to the same cross section
  with an upper cut of 1 TeV on the $HHH$ invariant mass (red area).
  Right: same plot zoomed around the SM configuration. 
  The integrated luminosity assumed is about 20 times larger than the reference luminosity in table~\ref{tab:sL}.
 }
   \label{fig:lim3}
\end{figure}

\begin{figure}
  \includegraphics[width=0.45\textwidth]{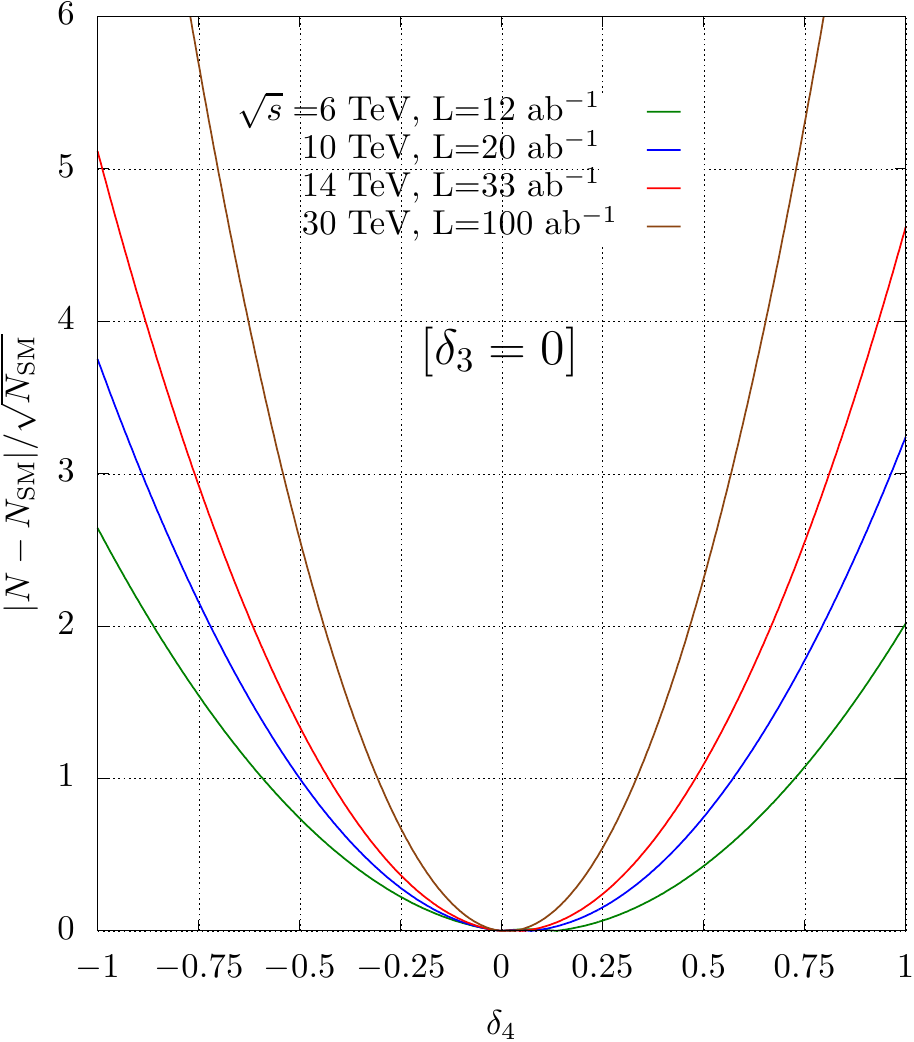}
  \hskip 1 cm
  \includegraphics[width=0.45\textwidth]{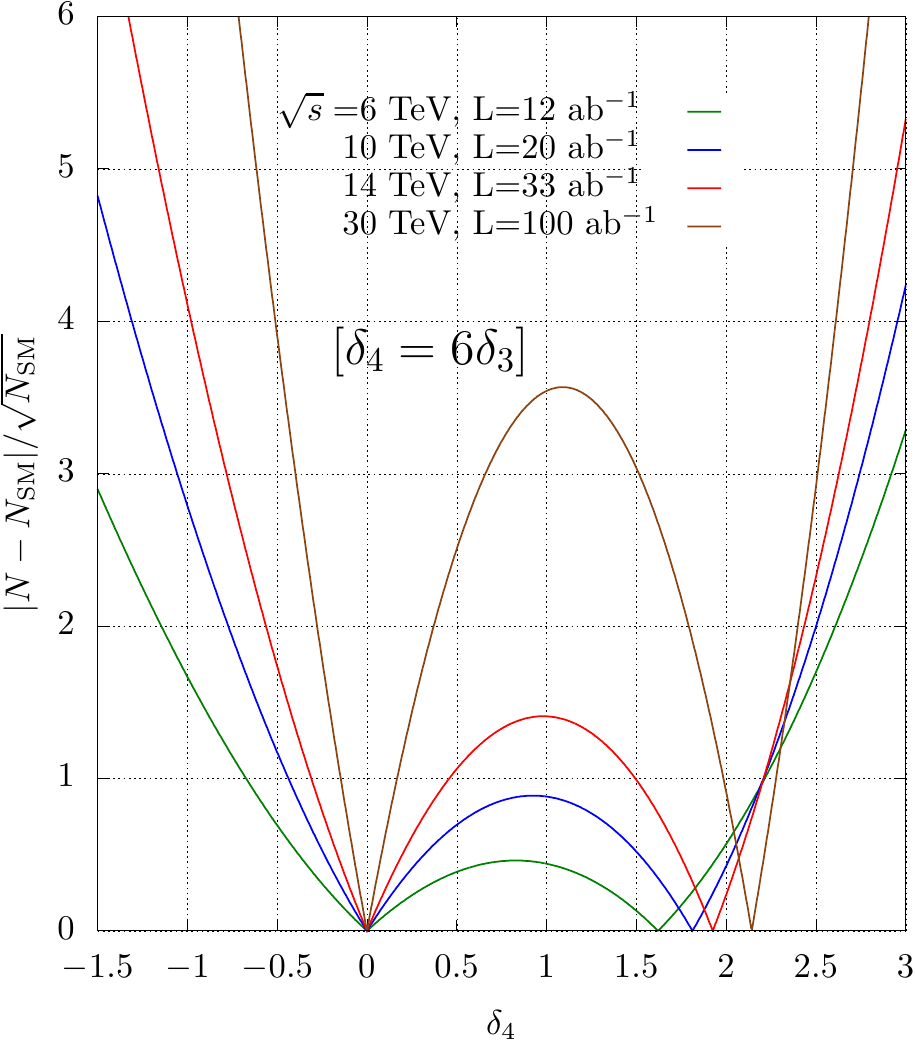}
   \caption{Sensitivity to the quartic Higgs self-coupling in terms of  standard deviations 
  $|N-N_{\rm SM}|/\sqrt{N_{\rm SM}}$ with respect to  the SM configuration,
  where the event numbers $N$ 
  refer  to $\sigma(\mu^+ \mu^- \to H H H \nu \overline{\nu})$,  for $M_{\bar\nu\nu}\gtrsim 150$GeV,  for  $\delta_3=0$ (left), and  $\delta_4=6 \delta_3$ (right). Results are obtained considering deviations from
  the inclusive cross sections only.}
\label{fig:sigma}
\end{figure}

\begin{figure}
  \includegraphics[width=0.45\textwidth]{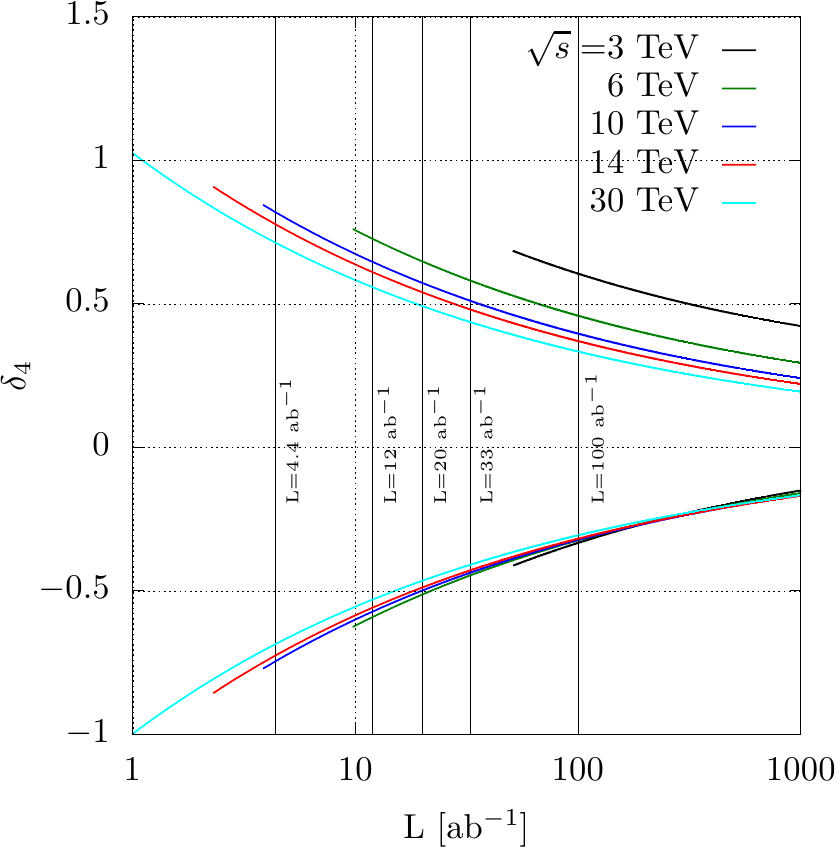}
  \hskip 1 cm
  \includegraphics[width=0.45\textwidth]{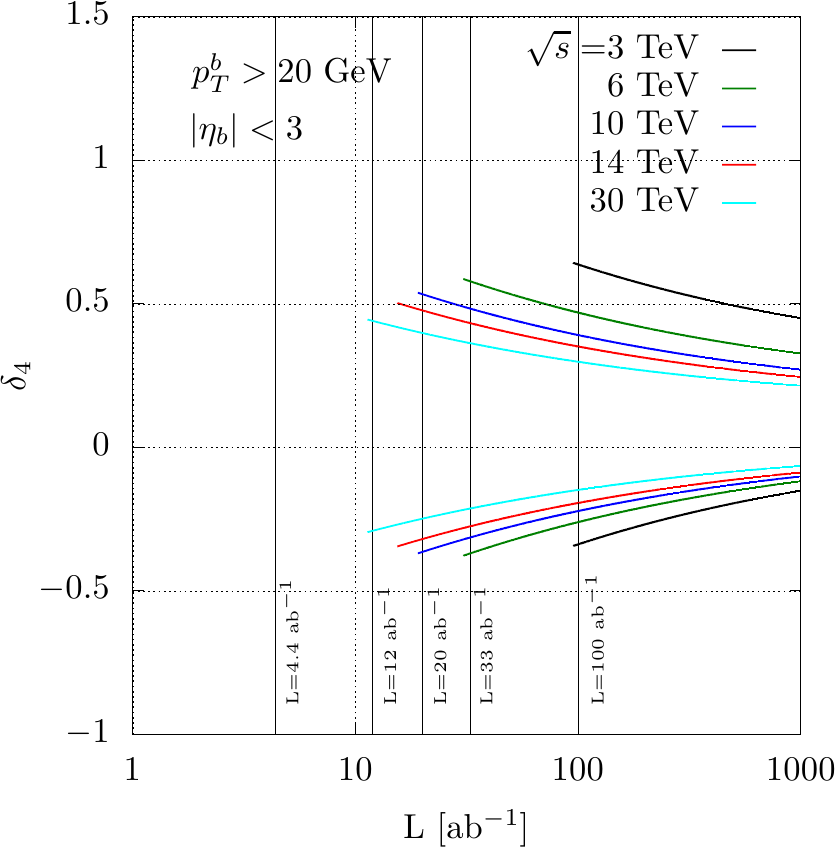}
   \caption{   \label{fig:d4vsL} One-sigma constraints on $\delta_4$ assuming $\delta_3=$0, for the c.m. energies in table~\ref{tab:sL}, as a function of the integrated luminosity. The curves are
   obtained by requiring at least twenty signal events. The left plot corresponds to the inclusive setup,
   while in the right plot the Higgs decay products must have $p_T>$20~GeV and $|\eta|<$3.}

\end{figure}

\begin{figure}
  \includegraphics[width=0.45\textwidth]{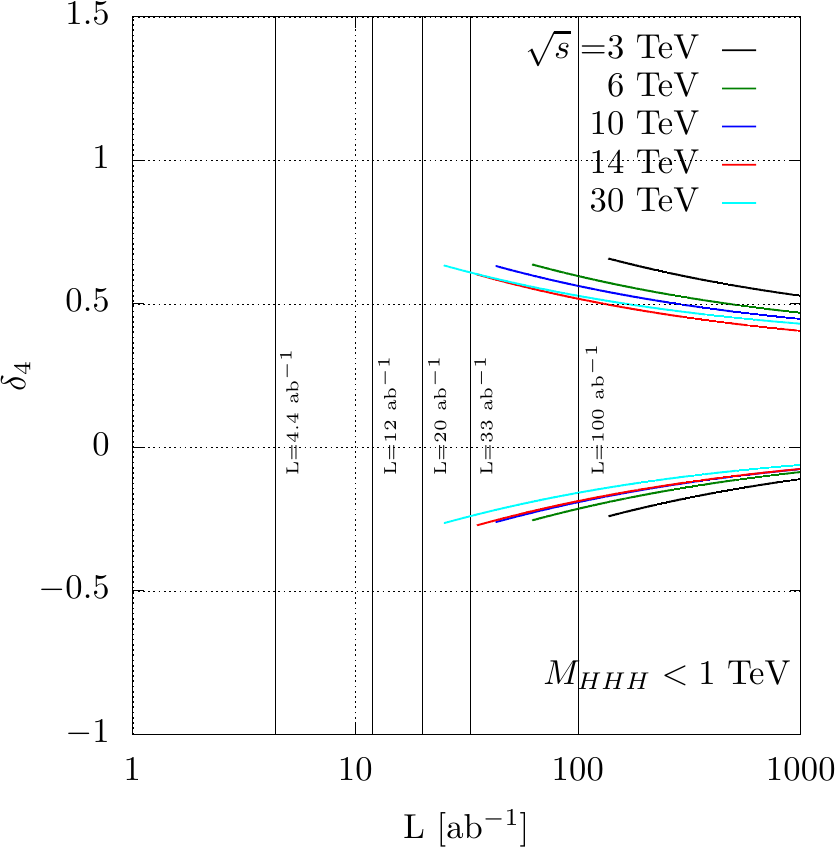}
  \hskip 1 cm
  \includegraphics[width=0.45\textwidth]{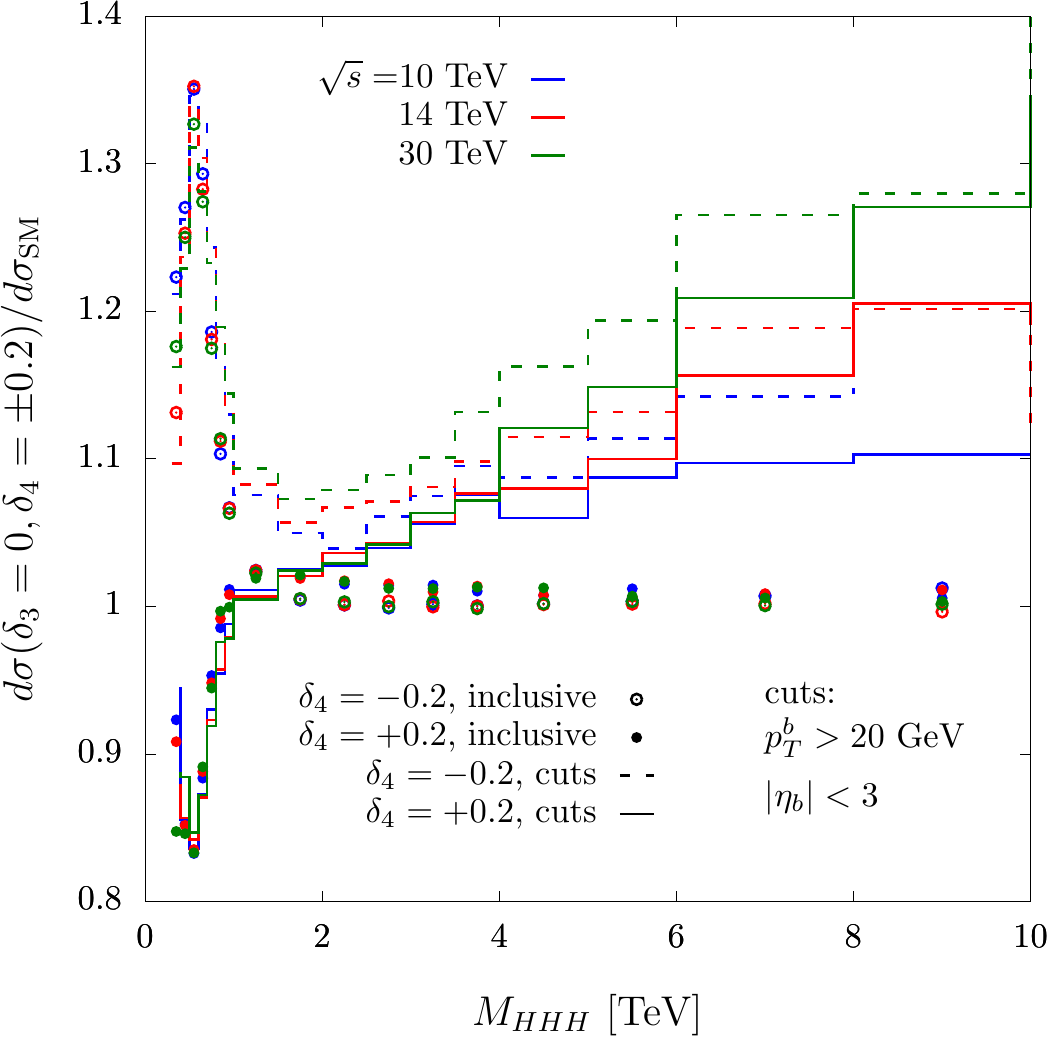}
  \caption{   \label{fig:d4vsL2} Left plot: one-sigma constraints on $\delta_4$ in the threshold region
    ($M_{HHH}<1$~TeV) assuming $\delta_3=$0, for the c.m. energies in table~\ref{tab:sL}, as a function of
    the integrated luminosity. The curves are obtained by requiring at least twenty signal events.
    Right plot: ratios of the invariant mass distributions of the three Higgs bosons for $\delta_4=\pm 0.2$
    ($\delta_3=0$) and in the SM. The dots correspond to the inclusive setup, while the solid/dashed
    curves are obtained by imposing the geometric acceptance cuts on the Higgs decay products.}

\end{figure}

\begin{figure}
  \includegraphics[width=0.45\textwidth]{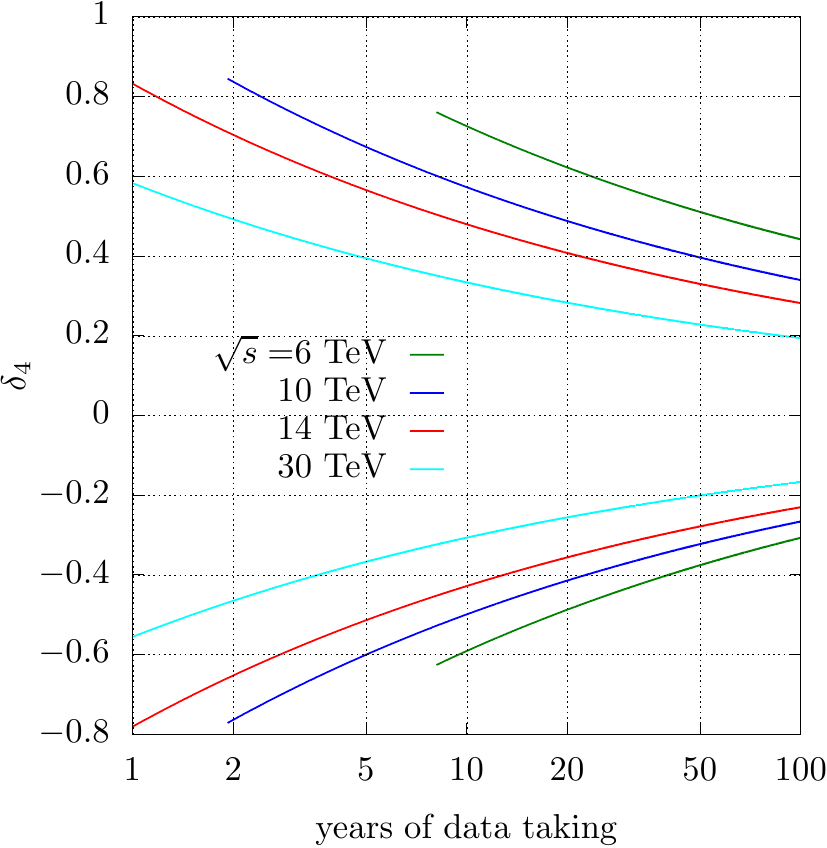}
  \hskip 1 cm
  \includegraphics[width=0.45\textwidth]{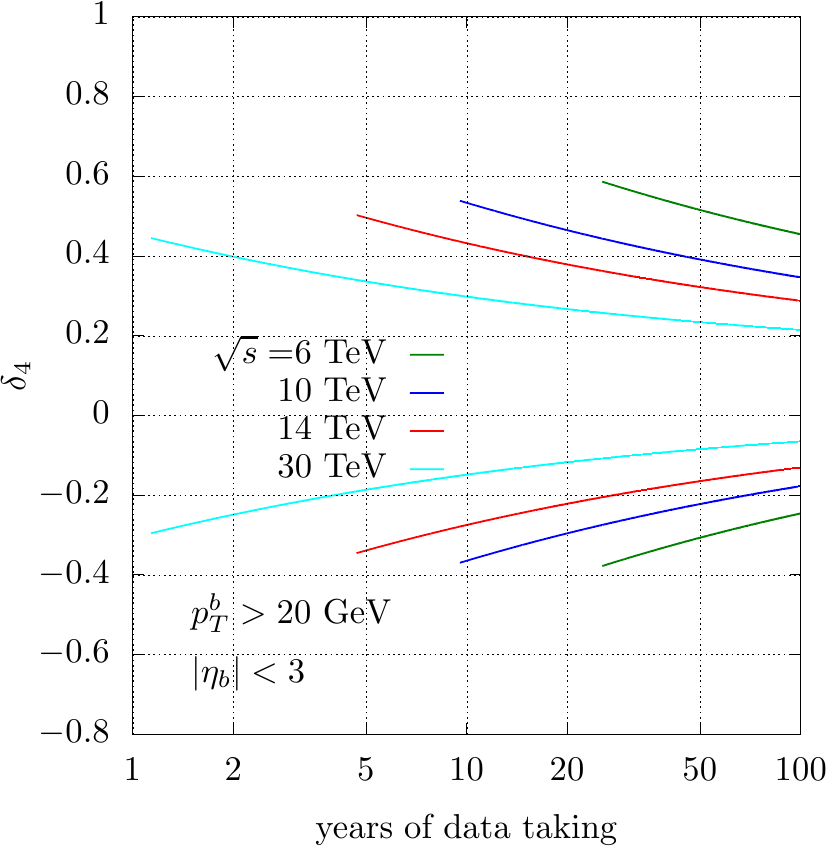}
  \caption{One-sigma constraints on $\delta_4$ assuming $\delta_3=$0 for the
    c.m. energies and the instantaneous luminosities in table~\ref{tab:sL} as a function of the
     number of years of data taking. The curves are obtained by requiring at least twenty signal events.
     The left plot corresponds to the inclusive setup, while in the right plot the Higgs decay products
     must have $p_T>$20~GeV and $|\eta|<$3. The curves corresponding to the 3~TeV setup fall outside the
     time interval under consideration}
\label{fig:d4vsY}
\end{figure}

\begin{figure}
  \includegraphics[width=0.45\textwidth]{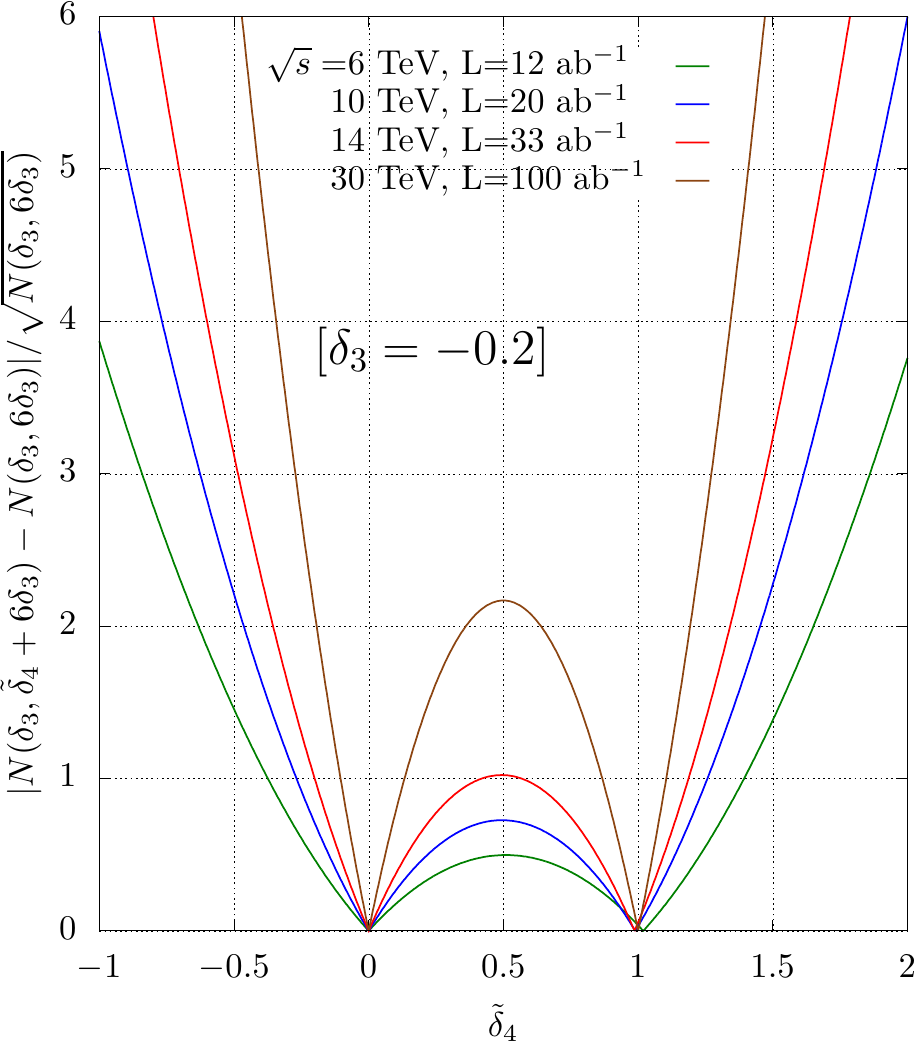}
  \hskip 1 cm
  \includegraphics[width=0.45\textwidth]{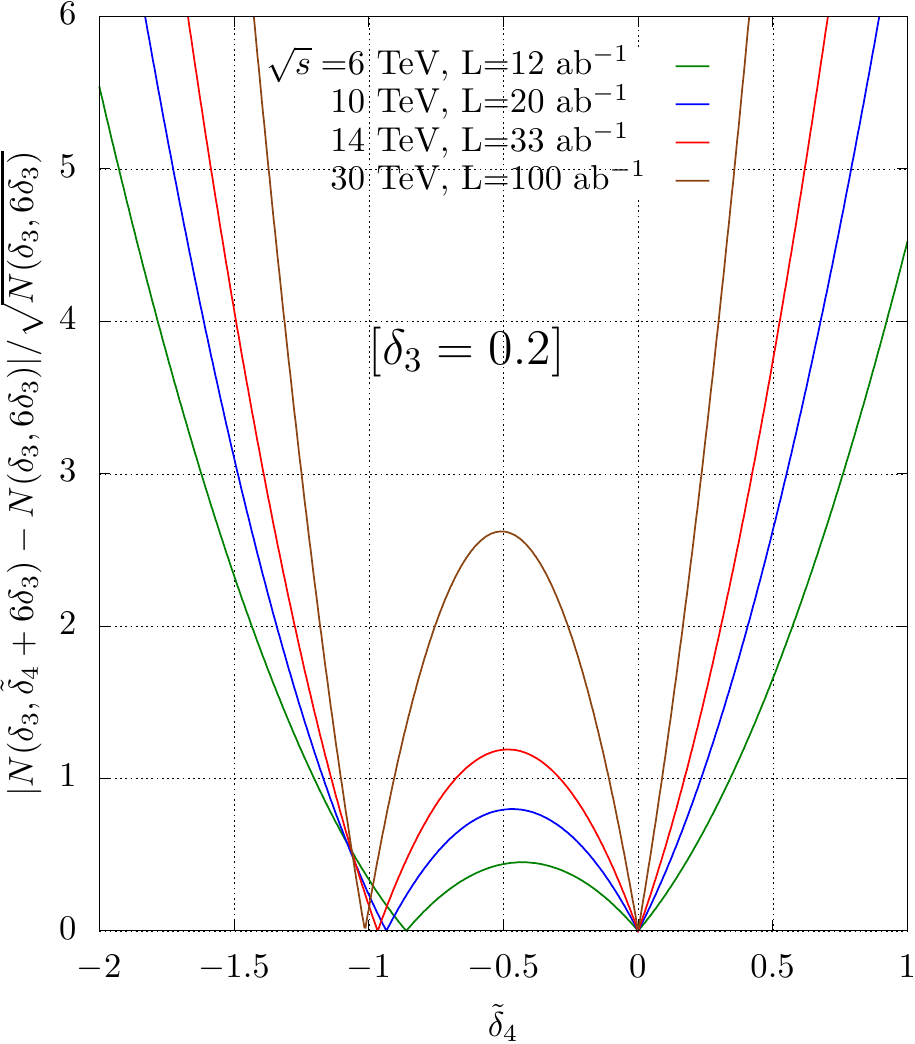}\\
  \includegraphics[width=0.45\textwidth]{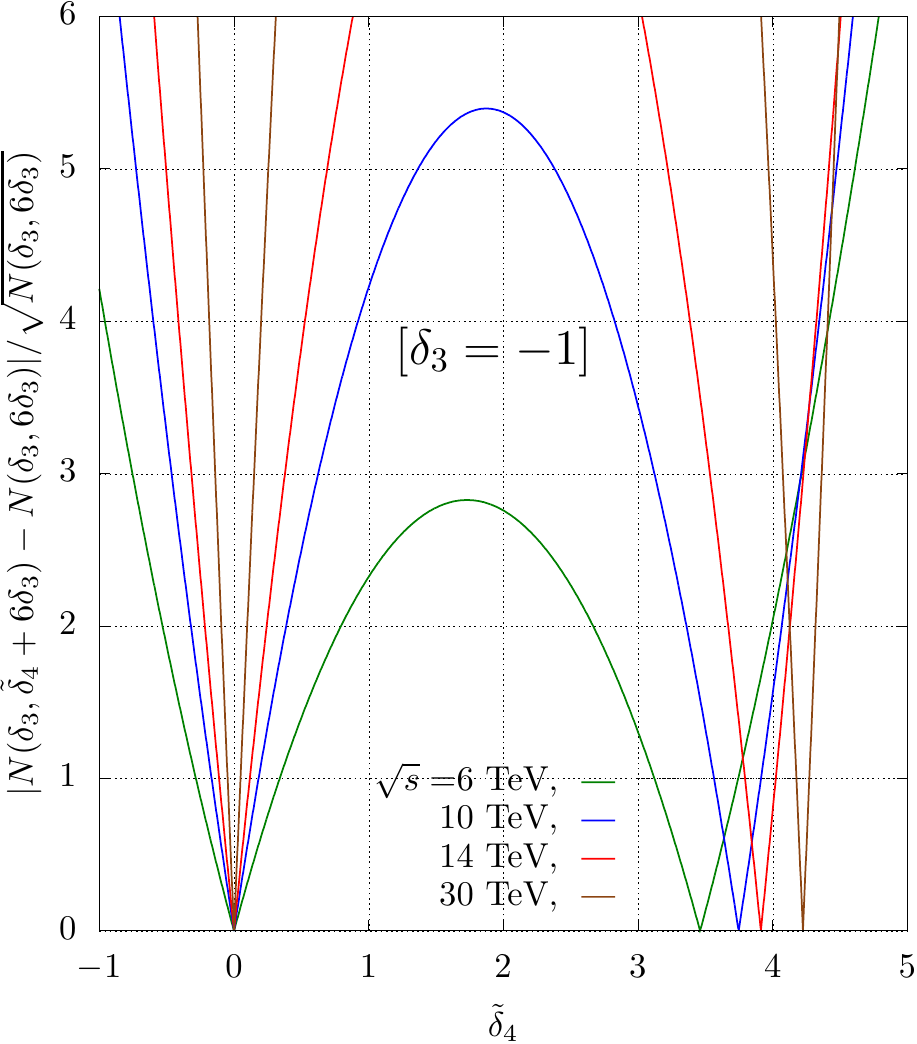}
  \hskip 1 cm
  \includegraphics[width=0.45\textwidth]{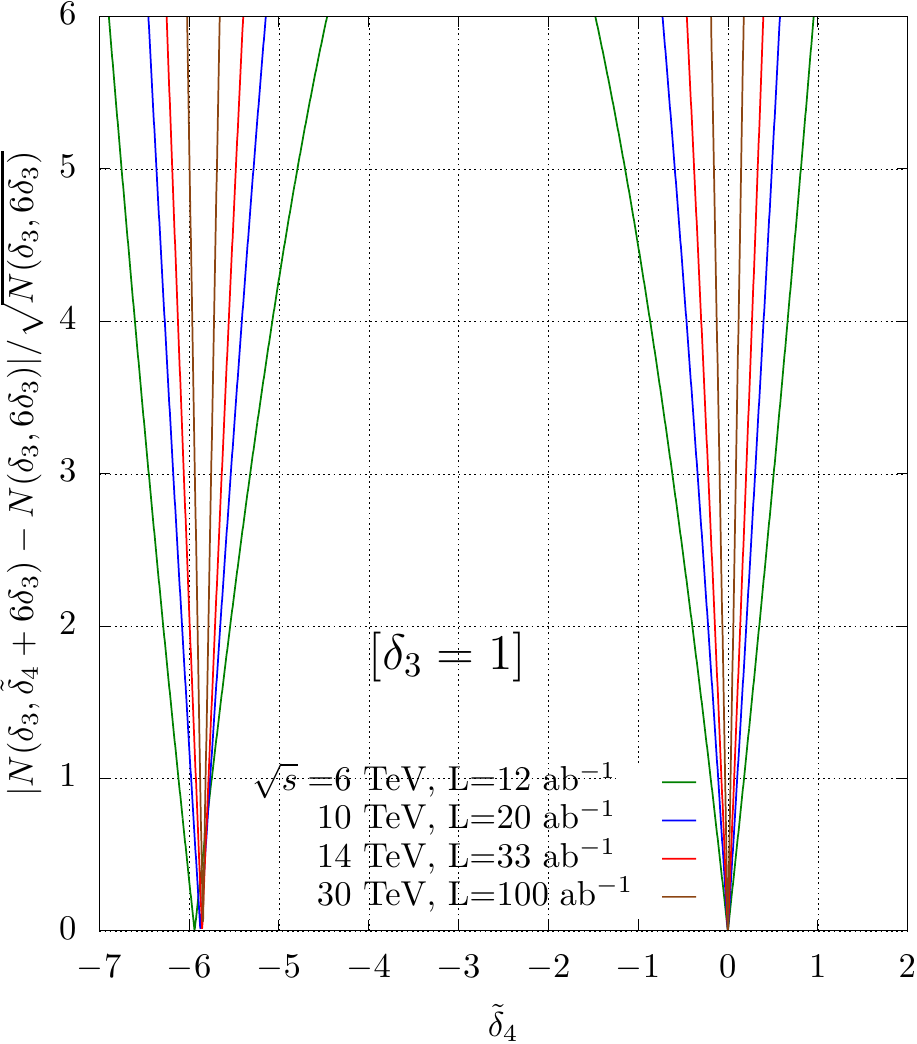}
   \caption{Sensitivity to $\tilde \delta_4=\delta_4-6 \delta_3$ in terms of  standard deviations 
  $|N(\delta_3,\tilde \delta_4+6 \delta_3)-N(\delta_3,6 \delta_3)|/\sqrt{N(\delta_3,6 \delta_3)}$ with respect to  the SMEFT configuration,
  where the event numbers $N$ 
  refer  to $\sigma(\mu^+ \mu^- \to H H H \nu \overline{\nu})$,  for $M_{\bar\nu\nu}\gtrsim 150$GeV. We assume 
    $\delta_3=-0.2$ (left), and  $\delta_3=0.2$ (right) in the upper plots,
    and $\delta_3=-1$ (left), and  $\delta_3=1$ (right) in the lower plots. Results are obtained considering deviations from
  the inclusive cross sections.}
\label{fig:d4tilde}
\end{figure}

\begin{figure}
  \includegraphics[width=0.45\textwidth]{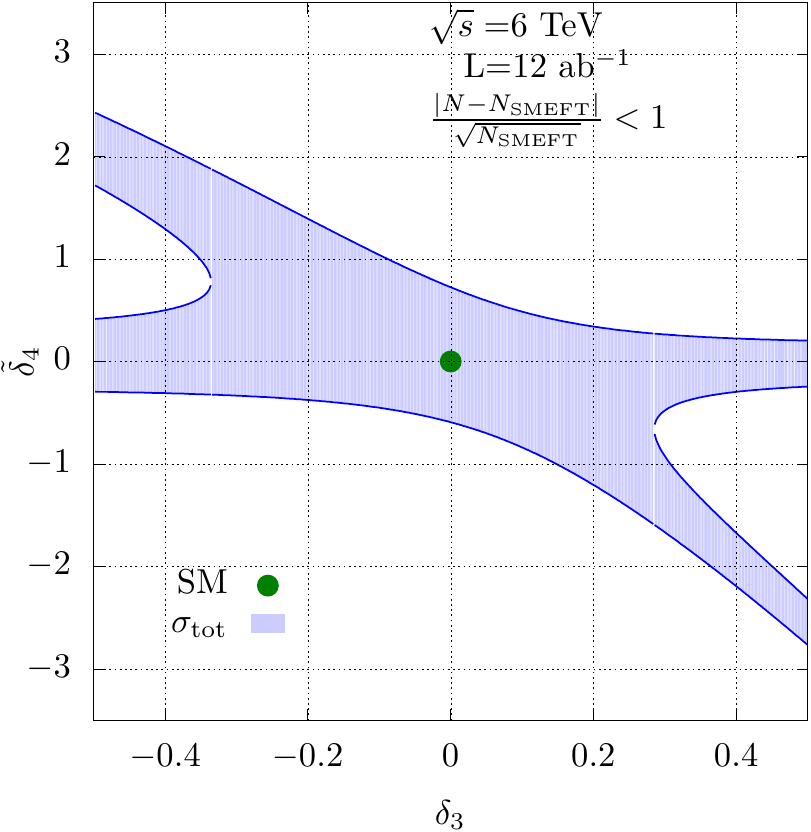}
  \hskip 1 cm
  \includegraphics[width=0.45\textwidth]{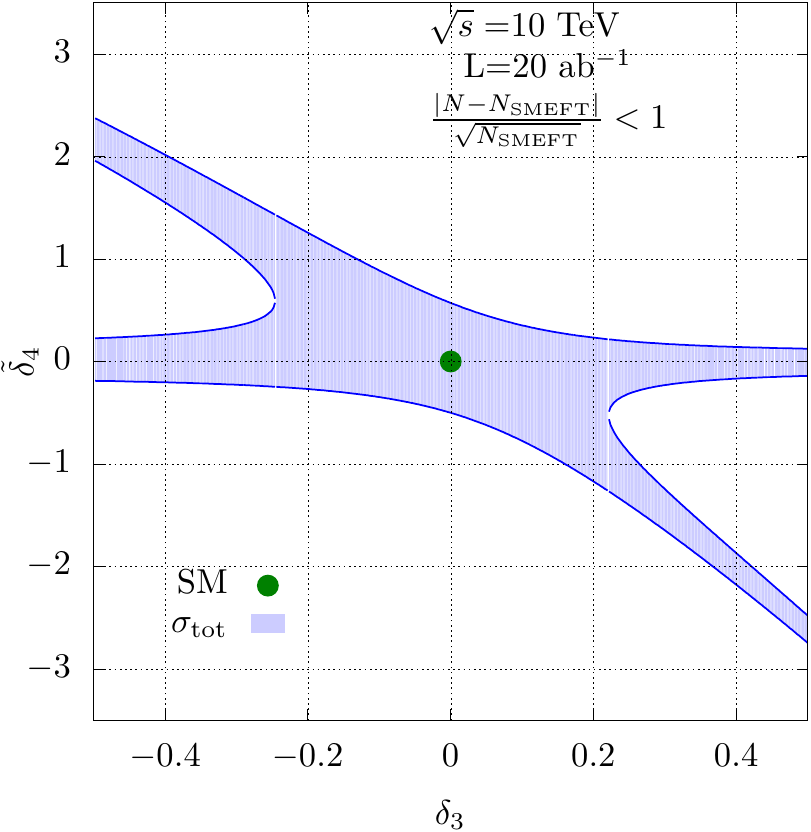}\\
  \includegraphics[width=0.45\textwidth]{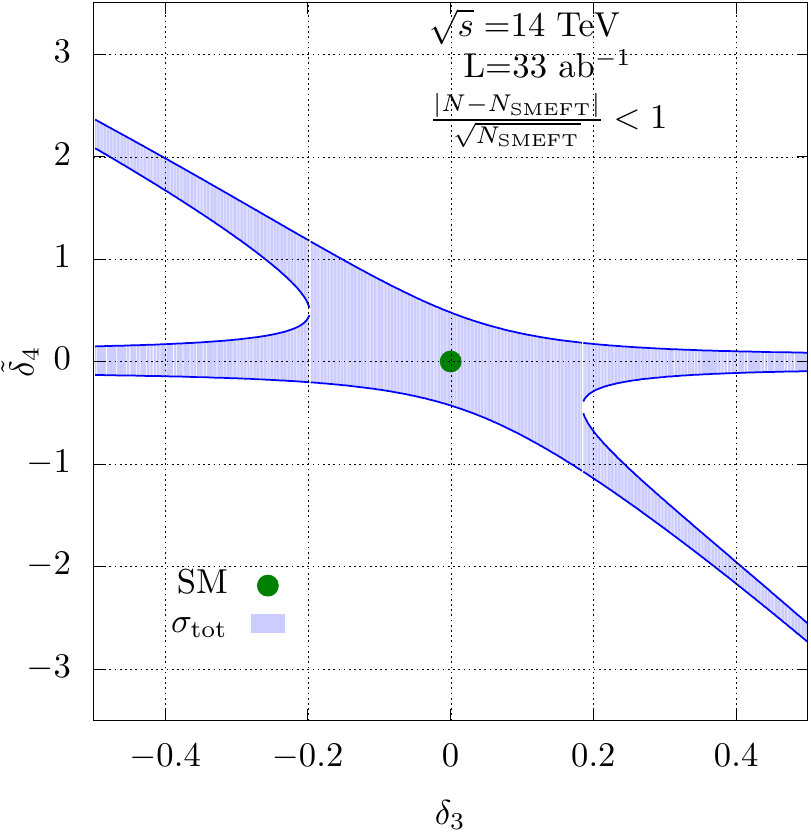}
  \hskip 1 cm
  \includegraphics[width=0.45\textwidth]{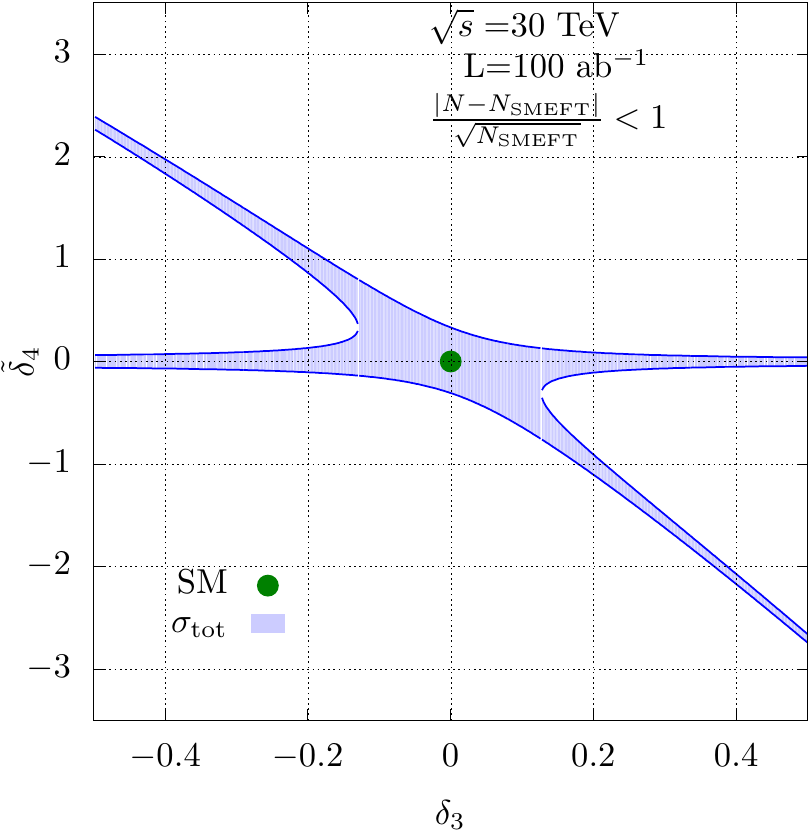}
  \caption{One-sigma exclusion plots in the ($\delta_3$, $\tilde{\delta}_4$) plane
    in terms of  standard deviations 
    $|N(\delta_3,\tilde \delta_4+6 \delta_3)-N(\delta_3,6 \delta_3)|/\sqrt{N(\delta_3,6 \delta_3)}$
    with respect to  the SMEFT configuration ($\tilde{\delta}_4=\delta_4-6\delta_3$),
    for the c.m. energy of 6 (top left), 10 (top right), 14 (bottom left), and 30~TeV (bottom right).
    The generation cut $M_{\bar\nu\nu}\gtrsim 150$GeV is applied, but no cut is imposed on the Higgs
    bosons or their decay products.}
\label{fig:d4tildeCT}
\end{figure}

\begin{table}
\begin{center}
  \begin{tabular}{|cc|ccc|}
    \hline
 & & \multicolumn{3}{c|}{Constraints  on $\delta_4$ (with $\delta_3=0$)}    \\
$\sqrt{s}$ (TeV)    &    Lumi (ab$^{-1}$)      &  x-sec only &  x-sec only &  threshold + $M_{HHH}>1$~TeV   \\
  & &   1 $\sigma$ & 2 $\sigma$ &  1 $\sigma$ \\
\hline
6                     & 12                                 & $[-0.60, 0.75]$& $[-0.90, 1.00]$ & $[-0.55, 0.85]$\\
10                    & 20                                 & $[-0.50, 0.55]$& $[-0.70, 0.80]$ & $[-0.45, 0.70]$\\
14                    & 33                                 & $[-0.45, 0.50]$& $[-0.60, 0.65]$ & $[-0.35, 0.55]$\\
30                    & 100                                & $[-0.30, 0.35]$& $[-0.45, 0.45]$ & $[-0.20, 0.40]$\\
3                     & 100                                & $[-0.35, 0.60]$& $[-0.50,0.80]$  & $[-0.45, 0.65]$\\
\hline
\end{tabular}
\caption{Summary of the constraints on the quartic deviations $\delta_4$, assuming $\delta_3=0$,
  for various muon collider energy/luminosity options, as obtained from the total expected cross sections (1$\sigma$ and 2$\sigma$ CL).
  The third column shows the bounds obtained from the combination of the constraints corresponding to the setups
  $M_{HHH}<1$~TeV and $M_{HHH}>1$~TeV.}
\label{tab:summary}
\end{center}
\end{table}

\begin{table}
\begin{center}
  \begin{tabular}{|cc|ccc|}
    \hline
 & & \multicolumn{3}{c|}{Constraints  on $\delta_4$ (with $\delta_3=0$)}    \\
$\sqrt{s}$ (TeV)    &    Lumi (ab$^{-1}$)      & \multicolumn{3}{c|}{ x-sec only, acceptance cuts}   \\
  & &   1 $\sigma$ & 2 $\sigma$ &  3 $\sigma$ \\
\hline
6                     & 12                                 & $[-0.50, 0.70]$& $[-0.74, 0.95]$ & $[-0.93, 1.15]$\\
10                    & 20                                 & $[-0.37, 0.54]$& $[-0.55, 0.72]$ & $[-0.69, 0.85]$\\
14                    & 33                                 & $[-0.28, 0.43]$& $[-0.42, 0.58]$ & $[-0.52, 0.68]$\\
30                    & 100                                & $[-0.15, 0.30]$& $[-0.24, 0.38]$ & $[-0.30, 0.45]$\\
3                     & 100                                & $[-0.34, 0.64]$& $[-0.53,0.82]$  & $[-0.67, 0.97]$\\
\hline
\end{tabular}
  \caption{Constraints on $\delta_4$ ($\delta_3=$0)
    for the c.m. energies and the instantaneous luminosities in table~\ref{tab:sL}
    once the geometric acceptance cuts $p_T>$20~GeV and $|\eta|<3$ are applied to the
    Higgs decay products. The bounds are obtained from the total expected cross sections
    for the process $\mu^+\mu^- \to HHH\nu\bar{\nu}$.
    The Higgs bosons are produced on-shell and decayed to $b\bar{b}$ pairs but no branching ratio is applied.}
\label{tab:summarycuts}
\end{center}
\end{table}

In figure~\ref{fig:2s}  we show the dependence on $\delta_4$ and $\delta_3$ of the total
cross section in two different bins, inclusive and for $M_{HHH}<1$ TeV, and for the A (left)
and B (right) scenarios (under the SM hypothesis), respectively.  
  In both scenarios, one finds that the dependence on the anomalous couplings is weaker at high energy
  in the inclusive setup, while at threshold it is basically the same for all the c.m. energies under consideration.
  This is a consequence of the fact that the  highest sensitivity is at threshold,
  while the region  $M_{HHH}>1$ TeV  is much less sensitive to $\delta_3$ and $\delta_4$ and
  gives a contribution to the total cross section that increases with $\sqrt{s}$. 
  This means that, as far as no further cuts are imposed on the Higgs
  decay products, increasing the energy brings an advantage mainly in the statistics.
The results corresponding to independent variations of $\delta_3$ and $\delta_4$ are shown
in figures~\ref{fig:lim1},~\ref{fig:lim2} and~\ref{fig:lim3}, where the red shaded areas correspond to
the  constraints obtained from threshold region, while the blue shaded areas  correspond to the full sample.
The plots on the right are blowups of the region close to the SM point $(0,0)$.
First, we note that as the energy increases, the blue areas tend to the shape of a ring in the plot range,
showing the relevance of the quadratic terms and the fact that bounds are obtained from upper as well as
lower limit in the number of events with respect to the SM expectations.
As expected from the arguments given above, the constraints improve as the energy/luminosity increase mostly for the blue areas. In addition, the linear flat direction in the case of same sign variations of $\delta_3$ and $\delta_4$ are resolved by using two different regions and the higher terms in the $c_i$ expansion. Figure~\ref{fig:lim3} indicates that low energy runs, around 3 TeV, yet with a  luminosity of 100 ab$^{-1}$ could provide a determination  in the range 
$-0.3 <\delta_4<0.6$ (with $\delta_3=0$).

Figure~\ref{fig:sigma} presents the sensitivity in terms of number of standard deviations for $\sqrt{s}=$6, 10, 14, and 30~TeV, 
assuming the integrated luminosities in table~\ref{tab:sL}. In the case $\delta_3=$0 (left plot), it shows that a three-sigma constraint
on $\delta_4$ falls in the interval [-1,1] for all the setups, with the only exception of $\sqrt{s}=$6~TeV, where the $\delta_4$
interval is slightly larger.

The constraints that can be obtained from the various energy/luminosity scenarios by using only information on the total cross section at 1$\sigma$ and $2\sigma$, and by combining events
in the regions $M_{HHH}<1$~TeV and $M_{HHH}>1$~TeV  ($1\sigma$), are summarized in table~\ref{tab:summary}. 

Figure~\ref{fig:d4vsL} (left plot) shows the one-sigma exclusion limits on $\delta_4$
(assuming $\delta_3=$0) for $\sqrt{s}=$3, 6, 10, 14, and 30~TeV,  varying the assumptions on the integrated luminosity.
The curves are obtained requiring at least twenty signal events. As expected, the sensitivity to $\delta_4$
improves when increasing the luminosity, rather moderately. 
Figure~\ref{fig:d4vsL} (right plot) presents the same limits on $\delta_4$, but after
imposing a minimum transverse momentum of 20~GeV and a maximum rapidity of 3 to the Higgs
decay products. The message from this plot is twofold. On the one hand, the restriction in phase-space makes more
difficult to reach the minimum number of events (20) required in the analysis (in particular,
it needs to change the assumption on~L for the 6~TeV setup from 12~ab$^{-1}$ to about 30~ab$^{-1}$).
On the other hand, interestingly, the plot also shows that the sensitivity to $\delta_4$ improves when
imposing a cut on the rapidity of the Higgs boson decay products. This is consequence
of the fact that, while the SM production of three Higgs bosons tends to be forward,
the effect of non-SM quartic Higgs couplings is mostly on central production, as one can naively expect
from the behaviour of the amplitudes in presence of four-point interactions.
Table~\ref{tab:summarycuts} summarizes the constraints on $\delta_4$ ($\delta_3=0$)
that can be obtained by imposing the geometric acceptance cuts $p_T>$20~GeV and $|\eta|<3$ to the
Higgs decay products for the c.m. energies and the instantaneous luminosities in table~\ref{tab:sL}.

The one-sigma exclusion limits on $\delta_4$ ($\delta_3=0$)
in the threshold region are shown in figure~\ref{fig:d4vsL2} (left plot). By comparing figures~\ref{fig:d4vsL2} and~\ref{fig:d4vsL},
one can notice that, imposing only the threshold
cut $M_{HHH} < 1$~TeV, the required luminosity increases by a factor of
about two and the sensitivity to positive values of $\delta_4$ decreases, 
compared to the limits obtained under the geometric acceptance cuts
on the Higgs decay products. These features can be explained by the fact that,
while at low energies most of the central events that have the largest
constraining power on $\delta_4$ fall in the threshold region, for
larger values of $\sqrt{s}$ there is a non negligible fraction of central
events with $M_{HHH}>1$~TeV. This can be read, for instance, from the
right plot in figure~\ref{fig:d4vsL2}, which shows the ratios of the $M_{HHH}$ distributions
computed for $\delta_4=\pm 0.2$ ($\delta_3=0$) and in the SM for several c.m.
energies both in the inclusive setup and after imposing a minimum
transverse momentum of 20~GeV and a maximum rapidity of~3 to the Higgs
decay products: while in the inclusive setup the high invariant-mass tail
is dominated by forward events and basically independent of $\delta_4$,
when the acceptance cuts are imposed the ratios become positive for
large invariant masses and increase with $M_{HHH}$. This happens for
both positive and negative values of $\delta_4$. However, for $\delta_4<0$
most of the sensitivity corresponds to the positive peak at low invariant
masses in figure~\ref{fig:d4vsL2}. From the comparison of figures~\ref{fig:d4vsL}
and~\ref{fig:d4vsL2}, we also notice that the sensitivity to $\delta_4$ improves
with the c.m. energy when the acceptance cuts are imposed, while
the constraints on positive values of $\delta_4$ are slightly more
stringent at 14~TeV compared to the ones at 30~TeV if only the cut
$M_{HHH}<1$~TeV is applied. A similar behaviour can be found in figure~\ref{fig:2s},
where the curve corresponding to $\sqrt{s}=14$~TeV in the threshold region
is above the one for $\sqrt{s}=30$~TeV for $\delta_4>0$.

The plots in figure~\ref{fig:d4vsY} are closely related to the ones in
figure~\ref{fig:d4vsL}, and present the one-sigma constraints on $\delta_4$
(when $\delta_3=$0) as a function of the number of years of data taking, both with   
and without cuts on the Higgs decay products, assuming
the instantaneous luminosities in table~\ref{tab:sL}. The curves
corresponding to the 3~TeV setup fall outside the time interval
under consideration. 

In the present discussion, the underlying assumption for the setup $\delta_3=0$ is that no deviations are previously measured from the SM triple Higgs self-coupling. However, in case the study of $HH$ production at the muon collider or at other machines will discover deviations from $\delta_3=0$,
it would be interesting to search for possible deviations of $\delta_4$ from its expected value in the SMEFT
($\tilde \delta_4=\delta_4-6 \delta_3$). As an example, the plots in figure~\ref{fig:d4tilde} show the sensitivity
to $\tilde \delta_4$ under the assumptions $\delta_3=\mp 0.2$ and $\delta_3=\mp 1$ .

The one-sigma exclusion plots in the ($\delta_3$, $\tilde{\delta}_4$) plane are presented in figure~\ref{fig:d4tildeCT}, 
for $\sqrt{s}=$6, 10, 14, and 30~TeV assuming the integrated luminosities in table~\ref{tab:sL}. The $\delta_3$ range in the plots 
is motivated by the one-sigma limits on this parameter that can be reached at other future colliders, spanning from
$\pm$50\% at HL-LHC and CEPC, to $\pm$10\% at CLIC and ILC at high energy, up to $\pm$5\% at FCC
(we refer to~\cite{DiMicco:2019ngk,deBlas:2019rxi} for a summary of the exclusion limits on $\delta_3$
at the proposed future machines). In the case of small deviations from the SM Higgs triple coupling,
to each value of $\delta_3$ corresponds an asymmetric interval in $\tilde{\delta}_4$ around $\tilde{\delta}_4=$0. Note that,  
for $\delta_3=$0, $\tilde{\delta}_4=\delta_4$ and the bounds on $\tilde{\delta}_4$ are the same as the ones
obtained in figures~\ref{fig:lim1} and~\ref{fig:lim2} for $\delta_4$. For larger values of $\delta_3$,
the allowed region in $\tilde{\delta}_4$ becomes very narrow and splits into two disjoint intervals, 
one centered around $\tilde{\delta}_4=$0 and the other one in the negative (positive) $\tilde{\delta}_4$
region for $\delta_3>$0 ($\delta_3<$0).


\section{Conclusions}
Dreaming about a muon collider as a future option to study fundamental interactions of elementary particles at the energy frontier is becoming a widespread reality in the high-energy community. Technical obstacles that were previously thought as insurmountable are  turned into formidable challenges worth to be investigated, wild expectations into ambitious goals at an increasing pace.
In keep with the progress in understanding what could be really achieved at the accelerator and detector level in a not-too-far future, theoretical and phenomenological investigations are mandatory to fully establish the physics reach of a very high energy lepton collider. 

In this work, we have considered one of the most important and challenging task ahead of us in the on-going exploration and verification of the standard model, i.e., the characterization of the Higgs potential at low energy. Many studies exist on the perspectives to measure the trilinear Higgs self-coupling at future hadron and (up to 3 TeV) lepton colliders and there is a general expectation that a precision at a few percent level could be reached at some point.  For this first exploration, we have therefore focused on the fourth derivative, the quartic self-coupling, whose determination is expected to be  extremely difficult at all foreseen colliders. 

We have considered in detail weak boson fusion production of three Higgs bosons, studying the sensitivity of total rates as well as of distributions on the Higgs boson self couplings.  We have found good sensitivity in the 
  threshold region, however, for the highest c.m. energy currently foreseen,
  the high energy tail provides most of the statistics and allows to improve the sensitivity to $\delta_4$ by restricting the acceptance to the central
  region.  We have then considered various possible scenarios attainable in different energy/luminosity configurations and determined the limits on the trilinear and quartic couplings in two motivated reference scenarios. Even though we adopted  (theoretical as well as experimental) simplifying assumptions, we reckon to have identified the most important features of this process. In particular, we have verified that even in the case where the forward regions become difficult to access due to machine backgrounds,  the sensitivity would not worsen as  mostly coming  from events central in the detector. This preliminary result  supports the hope that the possible worsening of the sensitivity from a more realistic analysis including, for instance,  backgrounds and  systematic uncertainties, might be offset by many possible improvements both in  analyses optimisation and in  detector  development.

Our results give a first indication that a leptonic collider at several TeV's of c.m. energy and with integrated luminosities of the order of a few  tens of attobarns, could provide enough events to allow a determination of the (SM) Higgs quartic  self-coupling with an accuracy in the tens of percent. For example, assuming $\lambda_3=\lambda_{SM}$,  and a (14 TeV/33~ab$^{-1}$) scenario, one could constrain $\lambda_4$ with a 50\% uncertainty at 1$\sigma$, i.e., significantly better than what is currently expected to be attainable at the FCC-hh with a similar luminosity. 

To finally assess the reach of a multi-TeV muon collider many more (and more detailed) studies will be necessary. This first work on the determination of the quartic self-coupling of the Higgs suggests that such studies are certainly worth to be undertaken.

\section*{Acknowledgments}

We thank Nadia Pastrone, Jean-Pierre Delahaye, Thorsten Ohl, Donatella Lucchesi, and Andrea Wulzer
for useful discussions. This work has received funding from the European Union's Horizon 2020 research
and innovation programme as part of the Marie Sk\l odowska-Curie Innovative Training Network MCnetITN3
(grant agreement no. 722104). The work of M.~C. has been supported by the Investissements d'avenir, Labex
ENIGMASS.

\end{document}